\documentclass[review]{elsarticle}
\usepackage{todonotes}
\usepackage{comment}
\usepackage{lineno}
\modulolinenumbers[5]
\usepackage[ruled,linesnumbered]{algorithm2e}
\usepackage{ amssymb }
\usepackage{amsfonts}
\usepackage{wrapfig}

  \usepackage[caption=false,font=footnotesize,labelfont=sf,textfont=sf]{subfig}
\usepackage[utf8]{inputenc}

\usepackage[T1]{fontenc}
\usepackage{lmodern}

\usepackage[breaklinks=true]{hyperref}

\usepackage{tikz}

\usetikzlibrary{positioning,automata}

\newtheorem{lemma}{Lemma}
\newdefinition{remark}{Remark}
\newdefinition{define}{Definition}
\newdefinition{axiom}{Axiom}
\newproof{proof}{Proof}
\newproof{pot}{Proof of Theorem \ref{thm2}}

\journal{Information Sciences}

\begin{document}

\begin{frontmatter}

\title{The Connection between Process Complexity of Event Sequences and Models discovered by Process Mining}

\author[MB]{Adriano Augusto}
\author[HU]{Jan Mendling\corref{mycorrespondingauthor}}
\ead{jan.mendling@hu-berlin.de}
\author[WU]{Maxim Vidgof}
\author[WU]{Bastian Wurm}
\address[MB]{The University of Melbourne, 3010 Parkville, VIC, Australia}
\address[HU]{Humboldt-Universit{\"a}t zu Berlin, Unter den Linden 6, 10099 Berlin, Germany}
\address[WU]{Wirtschaftsuniversit{\"a}t Wien, Welthandelsplatz 1, 1020 Vienna, Austria}

\cortext[mycorrespondingauthor]{Corresponding author, authors are listed in alphabetical order.}

\begin{abstract}
Process mining is a research area focusing on the design of algorithms that can automatically provide insights into business processes. Among the most popular algorithms are those for automated process discovery, which have the ultimate goal to generate a process model that summarizes the behavior recorded in an event log. Past research had the aim to improve process discovery algorithms irrespective of the characteristics of the input log. In this paper, we take a step back and investigate the connection between measures capturing characteristics of the input event log and the quality of the discovered process models. To this end, we review the state-of-the-art process complexity measures, propose a new process complexity measure based on graph entropy, and analyze this set of complexity measures on an extensive collection of event logs and corresponding automatically discovered process models. Our analysis shows that many process complexity measures correlate with the quality of the discovered process models, demonstrating the potential of using complexity measures as predictors of process model quality. This finding is important for process mining research, as it highlights that not only algorithms, but also connections between input data and output quality should be studied.\end{abstract}
\begin{keyword}
Process complexity\sep Event sequence data \sep Event logs \sep Process mining \sep Graph entropy \sep Automated process discovery
\end{keyword}
\end{frontmatter}


\section{Introduction}\label{sec:intro}
\noindent Recent years have seen a drastic increase in the availability of event sequence data and corresponding techniques for analyzing business processes, healthcare pathways, or software development routines~\cite{van2007business,monroe2013temporal,goh2019actions}.
Process mining is a research area focusing on the design of techniques that can automatically provide insights into business processes by analyzing historic process execution data, known as event logs~\cite{ProcessMiningBook,DBLP:books/sp/DumasRMR18}.
In process mining research, various algorithms have been developed for automated process discovery. A recent study found a rich spectrum of 35 distinct groups of such algorithms scattered over more than 80 studies~\cite{augusto2018automated}. Much of this research on automated process discovery is motivated by the ambition to improve process discovery outputs, in terms of high precision and recall, while producing models that are simple and easy to understand
~\cite{augusto2018split}.

So far, this stream of research on improving process discovery algorithms has been largely driven by the implicit assumption that a better algorithm would generate a better process model, no matter the characteristics of the input event log. In fact, there are good reasons to question this narrow focus. First, research on computer experiments highlights that studying the effect of input data characteristics on output is an important objective in many research areas~\cite{santner2003design,mendling2021theory}. Second, research on classifiers demonstrates the benefits of selecting algorithms based on characteristics of the input data~\cite{jordan2002discriminative,ribeiro2014recommender,ribeiro2015method}. Third, in various application domains, establishing a solid understanding of how input characteristics influences output has led to fundamental algorithmic innovations~\cite{sanders2009algorithm}.
For these reasons, Kriegel et al. recommend factoring in the variation of input parameters over meaningful ranges when comparing algorithms~\cite{kriegel2017black}.

In this paper, we revisit the output quality of automated process discovery algorithms in light of these arguments. More specifically, we investigate the empirical connections between measures capturing process complexity in terms of process behavior recorded in an event log and the quality of the process models discovered from that event log, as well as which of these process complexity measures can serve as a suitable predictor of process discovery quality. To this end, we first review process complexity measures defined in prior research studies. We analyze their characteristics and categorize them according to what perspective of process complexity they capture. Then, noting that each measure relates to a different perspective, we propose a new measure of process complexity based on graph entropy, which can exhaustively capture process complexity from multiple perspectives. Lastly, we analyze the process complexity measures using a prototypical implementation and an evaluation over an extensive set of event logs and their corresponding automatically discovered process models. Our analysis shows that many process complexity measures (including our novel measure) correlate with the quality of the discovered process models. Our findings demonstrate the potential of using process complexity measures as predictors for the quality of process models discovered with state-of-the-art process discovery algorithms. Such a result is important for process mining research, as it highlights that not only algorithms, but also connections between input data and output quality should be studied.

The remainder of the paper is structured as follows. Section~\ref{sec:back} summarizes prior research on measuring process complexity and related studies. Section~\ref{sec:entropy} presents our process complexity measure, how it is calculated, and which properties it satisfies. Section~\ref{sec:eval} presents our evaluation and the main findings of this study. Section~\ref{sec:conclude} concludes the paper and draws ideas for future work.

\section{Background and Related Work}\label{sec:back}
\noindent In this section, we contextualize our study by discussing related work with a  focus on the quality of discovered process models, automated process discovery algorithms, and process complexity.

\subsection{Quality of Discovered Process Models}

\noindent Process discovery is the task that encompasses the understanding of a business process behavior and the representation of that behavior in the form of a process model~\cite{DBLP:books/sp/DumasRMR18}. Process mining research has developed various algorithms for automated process discovery. The algorithms analyze the information recorded in an input event log (i.e., the process execution data capturing the process behavior) and automatically generate a process model as an output. Several measures have been defined for assessing the quality of a discovered process model~\cite{ProcessMiningBook}. Precision, fitness, and simplicity are the most frequently used ones.

\begin{figure}[htb]
  \centering
  \includegraphics[width=\textwidth]{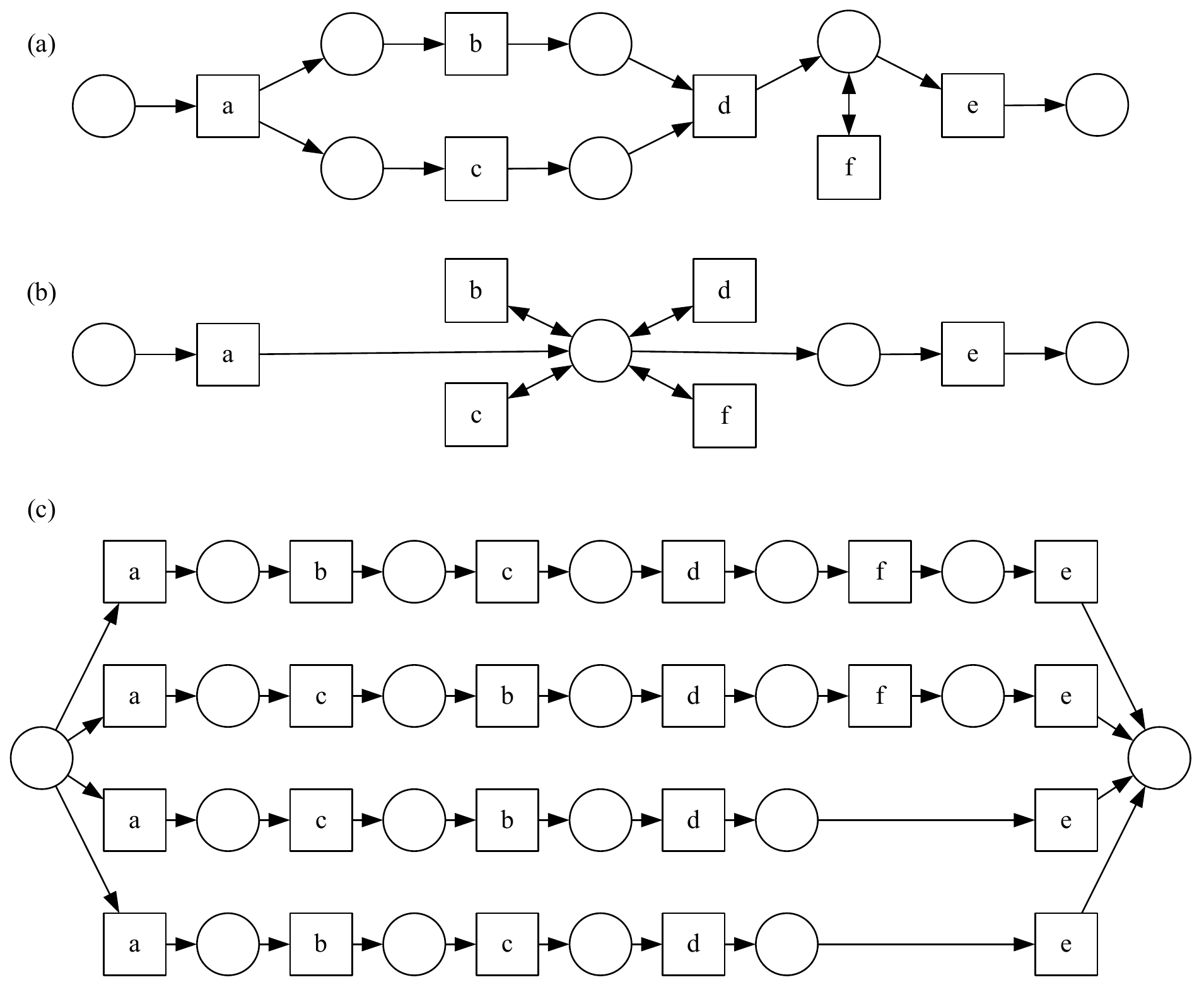}
  \caption{Considering the event log: $\langle a,b,c,d,f,e \rangle, \langle a,c,b,d,f,e\rangle, \langle a,c,b,d,e\rangle, \langle a,b,c,d,e\rangle$; three different algorithms may respectively discover the (Petri net) process models (a), (b) and (c).}
  \label{fig:criteria}
\end{figure}

Prior research studies have proposed several implementations of precision, fitness, and simplicity measures~\cite{syring11790evaluating,polyvyanyy2020monotone,cecconi2021measuring}. 
Let us assume that an event log is given, containing four sequences of events: $\langle a,b,c,d,f,e \rangle, \langle a,c,b,d,f,e\rangle, \langle a,c,b,d,e\rangle$ and $\langle a,b,c,d,e\rangle$. Let us also assume that three different automated process discovery algorithms have used this event log to construct the three Petri net process models shown in Figure~\ref{fig:criteria}. 

\emph{Precision} measures to which extent the process behavior (in terms of sequences of events) captured by the process model can be found in the original event log. It ranges between 0 and 1, where a value of 1 means that the process model can only generate sequences of events that are also contained in the event log. Model (c) is precise in this sense, Model (a) is less precise as it permits repetitions of the event $f$, and Model (b) is the least precise allowing any repetition and combination of the events $b,c,d,f$, as long as $a$ is the first and $e$ the last event.

\emph{Fitness} measures to which extent the behavior contained in the event log can be reproduced by the process model. Also fitness ranges between 0 and 1, where a value of 1 means that all the sequences of events contained in the event log can be reproduced by the process model. We observe that all Models (a), (b) and (c) have a fitness of 1, since all of them can reproduce the four sequences of events: $\langle a,b,c,d,f,e \rangle, \langle a,c,b,d,f,e\rangle, \langle a,c,b,d,e\rangle$ and $\langle a,b,c,d,e\rangle$. 

\emph{Simplicity} captures how easily a process model can be understood by a human. A model that is not simple is called \emph{complex}. Several complexity measures have been proposed in the literature~\cite{DBLP:books/sp/Mendling2008}. Some of them take into account the structure of a process model (e.g., model size), others the behavioral variability of a process model (e.g., control flow complexity). In our example in Figure~\ref{fig:criteria}, it is apparent that Model (c) is structurally more complex than Models (a) and (b), since it has more nodes and edges. On the other hand, Model (c) is behaviorally simpler than Model (a) and (b), since it allows for less variation. Accordingly, its control flow complexity is lower.

\subsection{Automated Process Discovery Algorithms}\label{sec:disalg}
\noindent Over the past decade, more than 80 research papers have proposed new algorithms for automated process discovery~\cite{augusto2018automated}. 
Often, the process models they produce significantly differ in terms of how they trade off precision, fitness, and simplicity~\cite{augusto2018automated}.
The latest benchmark study~\cite{augusto2018automated} compares and evaluates seven of the most effective state-of-the-art algorithms, namely,
$\alpha\$$~\cite{guo2015mining} (A\$),
Inductive Miner~\cite{leemans2014infrequent} (IM),
Evolutionary Tree Miner~\cite{BuijsDV12} (ETM),
Fodina~\cite{vanden2017fodina} (FO),
Structured Heuristic Miner 6.0~\cite{augusto2017automated} (SHM),
Split Miner~\cite{augusto2018split} (SM),
Hybrid ILP Miner~\cite{van2015ilp} (HILP).
Although there was no algorithm found to clearly dominate the others, 
IM, ETM, and SM turned out to be most reliable in terms of discovering 
fitting, precise, and simple process models across the whole benchmark dataset of 24 real-world event logs. However, they also showed a substantial variance of performance.
HILP and A\$ often discovered unsound models (i.e., containing behavioral errors, such as deadlocks) and rarely produced highly fitting, precise, or simple process models. While FO and SHM often produced accurate models, these were usually highly complex and difficult to interpret.

Even though the study by Augusto et al.~\cite{augusto2018automated} does not discuss this aspect, their results suggest a connection between the input event log features and the quality of the automatically discovered process models. 
Also other works have tried to select the most suitable discovery algorithm based on event log characteristics~\cite{ribeiro2014recommender,ribeiro2015method}, but without studying the connection between log complexity and model quality.
For this reason, we hypothesize that 
characteristics of the event log influence the quality of discovered process models. 

\subsection{Process Complexity as a Factor of Process Discovery Quality}
\noindent It is a challenge for discovery algorithms to generate process models that are easy to understand. Often, the generated models are overly complex. 
These complex models are called ``spaghetti models''. Van der Aalst emphasizes that ``spaghetti-like structures are not caused by the discovery algorithm but by the variability of the process''~\cite{van2012process}. In line with this observation, several proposals have been made for pre-processing event logs
independent of the discovery algorithm applied. These proposals build on clustering, supervised sequence labeling, sequential patterns, or text matching~\cite{cheng2015process,suriadi2017event,diba2020extraction}. They highlight the potential to improve process discovery outputs by modifying characteristics of the event log data as an input.

To assess the empirical connection between log complexity and process discovery quality, we require a specific measure that can adequately quantify process log complexity. Several evaluations of process mining algorithms have reported basic measures. There have been studies that focus on process model complexity~\cite{DBLP:books/sp/Mendling2008,mendling2010seven,sanchez2012quality,mendling2012thresholds}. 
The few studies that consider process complexity more explicitly stem from computer science, organization science, and management science. The corresponding measures are categorized in Table~\ref{tab:measures} and described next.

\begingroup
\begin{table}[tbp]
\makebox[\textwidth][c]{
\centering
{\scriptsize{
    \begin{tabular}{r|l|l|l}
    \hline
    \multicolumn{1}{l|}{Category} & Measure & Label & Ref. \\
    \hline
    \multicolumn{1}{l|}{Size} & Number of Events & magnitude & \cite{christianguenther-phd} \\
          & Number of Event Types & variety & \cite{christianguenther-phd} \\
          & Number of Sequences & support & \cite{christianguenther-phd} \\
          & Minimum, Average, Maximum Sequence Length & TL-min, TL-avg, TL-max & \cite{ProcessMiningBook} \\
          & Average Time Difference between Consecutive Events & (time) granularity &\cite{christianguenther-phd} \\
    \hline
    \multicolumn{1}{l|}{Variation} & Number of Acyclic Paths in Transition Matrix & LOD & \cite{DBLP:journals/misq/PentlandLKH20} \\
          & Number of Ties in Transition Matrix & t-comp & \cite{haerem2015task} \\
          & Lempel-Ziv Complexity & LZ & \cite{pentland2003conceptualizing} \\
          & Number and Percentage of Unique Sequences & DT(\#), DT(\%) & \cite{ProcessMiningBook} \\
          & Average Distinct Events per Sequence & structure & \cite{christianguenther-phd} \\
    \hline
    \multicolumn{1}{l|}{Distance} & Average Affinity & affinity & \cite{christianguenther-phd} \\
          & Deviation from Random & dev-random & \cite{pentland2003conceptualizing} \\
          & Average Edit Distance & avg-dist & \cite{pentland2003conceptualizing} \\
    \hline
    \end{tabular}%
    }}}
  \caption{Complexity Measures for Business Processes based on Event Logs}
  \label{tab:measures}%
\end{table}%
\endgroup
The first category includes \textit{size} measures. Various properties of an event log can be easily counted including the number of \textit{events}, \textit{sequences}, and \textit{event types}, the minimum, maximum, and average \textit{sequence length}, and the average and minimum \textit{time difference} between two events (proposed by G{\"u}nther~\cite[Ch.3]{christianguenther-phd}).

The second category contains measures related to the \textit{variation} of the process behavior recorded in the event log. Several of these measures take the transition matrix derived from the directly-follows relations observed in the event log as a starting point. Pentland~\cite{DBLP:journals/misq/PentlandLKH20} proposes the calculation of complexity as the number of \textit{acyclic paths} implied by the transition matrix derived from the event log. H{\ae}rem et al.~\cite{haerem2015task} use a slight variation based on what they call the number of \textit{ties}, which in essence is the count of
directly-follows relations observed in the event log. Also, Pentland's proposal~\cite{pentland2003conceptualizing} of measuring the number of operations of compressing the event log using the \textit{Lempel-Ziv} algorithm is a variation measure.
Finally, the (absolute and relative) number of \textit{distinct sequences}~\cite{ProcessMiningBook} and the average number of \textit{distinct events per sequence}~\cite{christianguenther-phd} also provide an indication of variation.

The third category refers to \textit{distance} measures. Several distance notions have been defined. Günther~\cite{christianguenther-phd} proposes the notion of \textit{affinity}, which is based on the overlapping directly-follows relations of two event sequences. His proposed complexity measure, namely \textit{average affinity}, is calculated as the mean of affinity over all pairs of event sequences~\cite[Ch.3]{christianguenther-phd}. This measure is closely related to the one proposed by Pentland~\cite{pentland2003conceptualizing} called \textit{deviation from random} of the transition matrix. Furthermore, Pentland also proposes a second distance measure based on the average \textit{edit distance} between event sequences based on notions of classical optimal matching~\cite{cornwell2015social}.

Each of these measures has its limitations and blind spots. We can easily identify cases where one measure indicates a difference while other measures are unaffected. We make the following observations on the relationship between two event logs $L_1$ and $L_2$:
\begin{description}
    \item[Observation O1:] Assume that $L_1$ and $L_2$ have the same size measures of events, event types and sequences. Let us consider the corner case where all the sequences of events recorded in $L_1$ are the same, while the sequences of events recorded in $L_2$ are all different. While size measures would not be able to capture any difference, variation and distance measures would detect them.
    \item[Observation O2:] Assume that $L_1$ and $L_2$ have the same variation measures in terms of the number of unique sequences. Still, $L_1$ and $L_2$ can be strikingly different in terms of size measures if the same sequences are repeated or not, and in terms of distance measures if the variants are very similar in one log and very different in the other.
    \item[Observation O3:] Assume that $L_1$ and $L_2$ have the same distance measures with each pair of sequences having, e.g., an edit distance of 1 on average. If $L_1$ includes each sequence of $L_2$ twice, the size measures will be substantially different. Also, the variation can be quite different if $L_1$ includes many sequences that are the same, while a few are very different, as compared to $L_2$ where all sequences have rather little distance.
\end{description}
These observations clearly show that there is no unique process complexity measure capable of capturing information regarding size, variation, and distance at once. In the next section, we propose a new measure that addresses this challenge. 

\section{Process Complexity based on Graph Entropy}\label{sec:entropy}
\noindent In this section, we propose a novel measurement of process complexity based on graph entropy.
More generally, entropy has been used for assessing the behavior of process representations and corresponding logs at the language level in~\cite{polyvyanyy2020monotone,polyvyanyy2020entropia}. This means that the measurements do not account for the distribution of variants of an event log.
Graph entropy is particularly suited as an underlying concept because it can capture size, variation, and distance in an integral way. To this end, we have to map an event log to a graph structure that acknowledges equivalences between sequences without introducing abstractions. In Section~\ref{3.1}, we define the notion of an extended prefix automaton. Section~\ref{3.2} defines graph entropy for extended prefix automata, proves monotonicity, and relates the proposed measure to Observations 1--3.

\subsection{Event Logs as Prefix Automata}\label{3.1}
\begin{figure}[htb]
  \centering
  \includegraphics[width=0.9\textwidth]{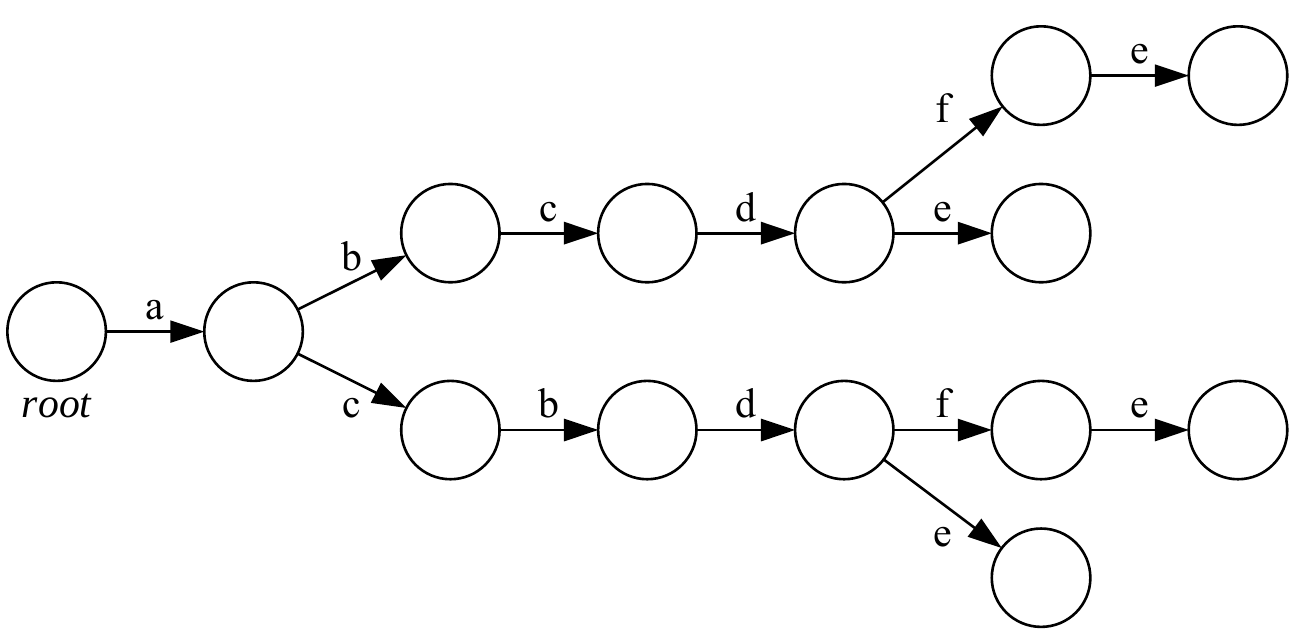}
  \caption{Prefix automaton derived from the previously discussed event log $\langle a,b,c,d,f,e \rangle, \langle a,c,b,d,f,e\rangle, \langle a,c,b,d,e\rangle, \langle a,b,c,d,e\rangle$.}
  \label{fig:prefix}
\end{figure}

\noindent In this section, we define the extended prefix automaton for an event log. The concept of a prefix automaton was introduced by Munoz-Gama and Carmona in~\cite{DBLP:conf/bpm/Munoz-GamaC10} based on concepts described by Van der Aalst et al.\ in~\cite{DBLP:journals/sosym/AalstRVDKG10}. The concept of a prefix automaton is particularly suited for our purpose of describing the complexity of an event log. Prefix automata describe sequences without loss of information and abstraction. They account for equivalent prefixes but do not introduce complexity that is not present in the event log. Figure~\ref{fig:prefix} illustrates the idea of a prefix automaton for the event log with the four sequences we discussed above. We observe that overlapping prefixes of the sequences lead to joint paths in the prefix automaton. All paths originate from the $root$. There are as many variants in this event log as there are nodes on the right-hand side without successors. These are four in our case.

We revisit basic notions of events, event sequences, and event logs upon which we will define the construction of the extended prefix automaton.
\begin{define}[Event, Event Sequence, Event Log~\cite{ProcessMiningBook}]
Let $E$ be a set of unique event identifiers. For each event $e \in E$, we define four attributes:
\begin{itemize}
    \item An $activity \in E \to A $ where $A$ is the set of activities. $activity(e)$ is an activity type that an event $e$ refers to.
    \item A timestamp $ts \in E \to TS$ where $TS$ is the set of timestamps. $ts(e)$ is the time when $e$ occurred.
    \item A case identifier $case \in E \to CID$ where $CID$ is the set of unique case identifiers. $case(e)$ is the case that the event $e$ is related to.
    \item A predecessor $pred \in E \to E~\cup \perp$ maps each event to a preceding event of the same case if such an event exists or to $\perp$ otherwise. $pred(e)$ is a predecessor of $e$ if they share the same case identifier $case(pred(e)) = case(e)$, if $ts(pred(e)) < ts(e)~\land~\nexists e' \in E: [ case(e') = case(e)~\land~ts(pred(e)) < ts(e') < ts(e)]$. $pred(e) = \perp $ if $\nexists e' \in E: [ case(e') = case(e)~\land~ts(e') < ts(e)]$.
\end{itemize}

A plain event log $L_{plain} \in E^*$ is a finite sequence of events (with events potentially relating to different cases). A plain event log is ordered by event timestamps and not by cases.
\end{define}
The connection between an event log $L$ and a corresponding plain event log $L_{plain}$ is trivial. $L_{plain}$ is a concatenation of all $e \in \bigcup\limits_{\sigma \in L} \sigma$. In the opposite direction, a trace $\sigma_{cid} = \langle e_1, \dots, e_n \rangle~|~ e_i \in E \land case(e_i) = cid \land ts(e_i) < ts(e_{i+1})$ for all $1 \leq i \leq n-1 $, and a log $L = \bigcup\limits_{e \in E} \sigma_{case(e)}$.

\cite{DBLP:conf/bpm/Munoz-GamaC10} defines a prefix automaton $PA=(S, T, A, s_0)$ with $S$ being a set of states, $A$ the set of activities, $T \subseteq S \times A \times S$ set of transitions and $s_0$ the initial state.
Based on this, we introduce the concept of an extended prefix automaton.
Compared to the prefix automaton $TS$ in~\cite{DBLP:conf/bpm/Munoz-GamaC10}, our automaton is extended in two ways. First, we define the $seq$ function that maps each state $s \in S$ to a set of events $seq(s) \subseteq E$ having the same prefix as the state itself. Second, we define a partitioning function $C$ that splits the extended prefix automaton $EPA$ into $ 0 \leq k \leq |L|$ partitions. Note that $|L|$ here refers to the number of traces in an event log $L$. We discuss partitioning in more detail below.

\begin{define}[Extended prefix automaton]
$EPA = (S_+, T, A, C, seq, root)$
\begin{itemize}
    \item $S_+=S\cup \{root\}$ is a set of states
    \item $A$ is a set of activities
    \item $T \subseteq S_+ \times A \times S$ is a set of transitions. Note that the \textit{root} has no incoming transitions.
    \item $C \in S_+ \to \mathbb{N}_0\cup \{\perp\}$ a partitioning function, defining for each state $s \in S_+$ the partition to which it belongs. $C(s)$ refers to a partition the state $s$ belongs to. A state can only belong to one partition. We write $C(root) = \perp$ because the root node does not belong to any partition.
    \item $seq \in S \to \wp(E)$ maps each state to events having the same prefix as the state. Note that every event in the log $L$ (or $L_{plain}$) corresponds to one and only one state: $\forall e \in L~ \exists s \in S: e \in seq(s)$ and $\forall s, s' \in S: s \neq s' \Rightarrow seq(s) \cap seq(s') = \emptyset$.
    \item $root \in S_+$ is the entry state of the automaton and corresponds to an empty prefix. (same as $s_0$ in~\cite{DBLP:conf/bpm/Munoz-GamaC10})
\end{itemize}
\end{define}
\begin{figure}[htb]
  \centering
  \includegraphics[width=0.9\textwidth]{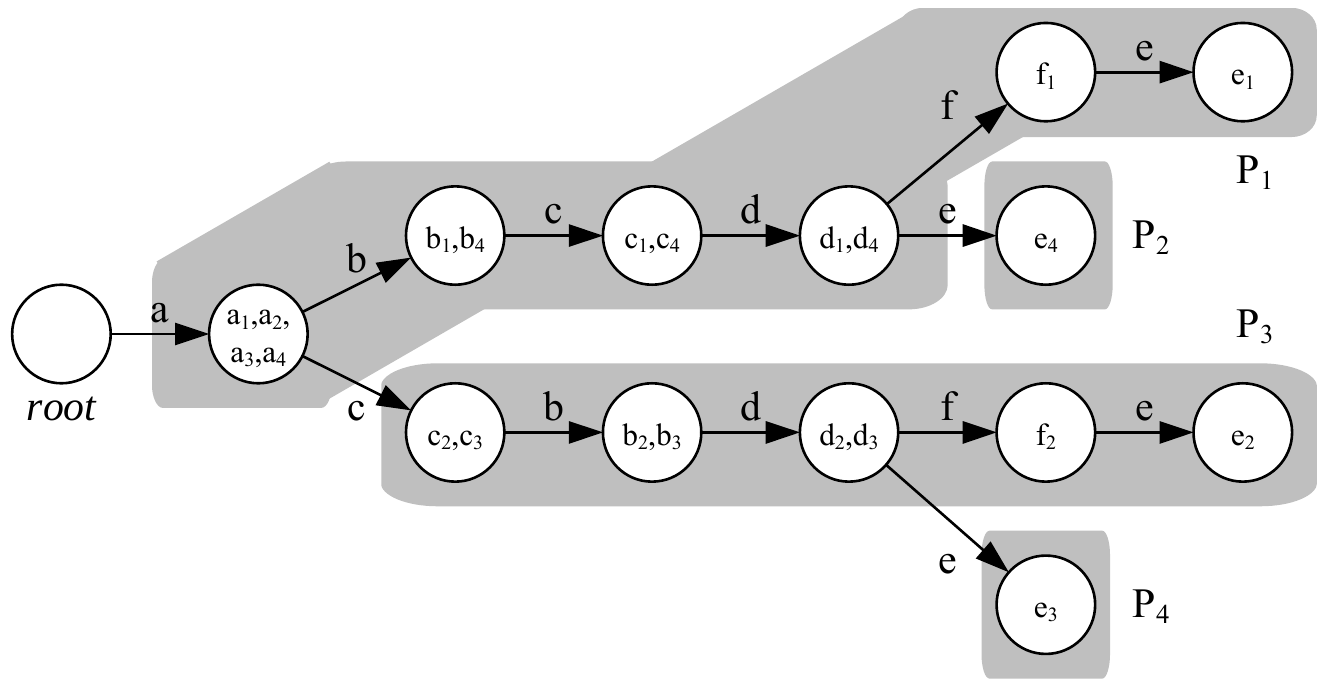}
  \caption{Extended prefix automaton with partitions derived from the event log $\langle a,b,c,d,f,e \rangle, \langle a,c,b,d,f,e\rangle, \langle a,c,b,d,e\rangle, \langle a,b,c,d,e\rangle$.}
  \label{fig:partition}
\end{figure}

Although the concept of accepting states is not mentioned in~\cite{DBLP:conf/bpm/Munoz-GamaC10} and mentioned only informally in~\cite{DBLP:journals/sosym/AalstRVDKG10}, we want to state explicitly here that in an extended prefix automaton all states are accepting states.\footnote{It is an alternative to introduce sink nodes similar to the root node as the only accepting states. We do not consider this alternative here, because it does not allow the incremental construction of the prefix automaton while events of running cases are continuously added.} This means that every trace $\sigma \in L$ can be replayed ($recall = 1$), but the extended prefix automaton can also produce shorter sequences that were not observed in the log $L$ ($precision \leq 1$). Note that this is no harm to our ambition of obtaining a representation of the event log with 100\% precision and recall when we store observed event sequences using the extended prefix automation.

Consider our previous example sequences and Figure~\ref{fig:partition}. The example illustrates how events of the event log are mapped to states with the help of the function $seq(s)$.
All sequences start with activity $a$, so one state $s^1_1$ suffices to capture all observed behavior up to that point. 
The second event in sequences 1 and 4 is $b$, which is captured by the state $s^1_2$ such that $C(s^1_2)=1$ and $seq(s^1_2)=\{b_1,b_4\}$. In sequences 2 and 3, the second event is $c$, which requires the extended prefix automaton to branch and introduce another state $s^3_2$ with $seq(s^3_2)=\{c_2,c_3\}$ in a new partition $P_3$.
Each time such branching occurs, a new partition is introduced, and the final number of partitions equals the number of observed process variants.

For its construction, we use the corresponding plain event log $L_{plain}$. Algorithm~\Ref{algorithm:building} shows how the extended prefix automaton is constructed. The algorithm iterates over the events in the log $L_{plain}$ and uses the variables $last_{AT}$, $pred_{AT}$, $current_{AT}$, and $current_c$.
The variable $last_{AT}$ is a mapping used to store the latest corresponding activity type that is added to the automaton for every case ID. While it is possible to search for it at every iteration, such a mapping increases the efficiency of the algorithm. $pred_{AT}$ is used to store the activity type of the current event's predecessor, while $current_{AT}$ stores the activity type of the event in question. $current_c$ stores the partition number of the current activity type.

\begin{algorithm}[p]
\scriptsize{
	\caption{Constructing the Extended Prefix Automaton}
	\label{algorithm:building}
	\SetAlgoLined
	\KwIn{Plain event log $L_{plain}$. }
	\KwResult{Extended Prefix Automaton $EPA$}
	Build $EPA = (S_+,T,A,C,seq,root)$ such that $S_+=\{root\}, T=A=seq=\emptyset, C(root) = 0$\;
	$last_{AT} \leftarrow \emptyset$\;
	\ForAll{event $e \in L_{plain} $}{
	    \eIf{$pred(e) \neq \perp$}{
	        $pred_{AT}\leftarrow last_{AT}(case(e)) $\;
	    }{
	        $pred_{AT}\leftarrow root$\;
	    }
	
	    $current_{AT} \leftarrow \emptyset$\;
	    \eIf{$\exists s | (pred_{AT}, activity(e), s) \in T$}{
	        $current_{AT} \leftarrow s$\;
	    }{
	        \eIf{$\exists a,s | (pred_{AT}, a, s) \in T$}{
	            $current_c \leftarrow max(C) + 1$\;
	        }{
	            \eIf{$pred_{AT}= root$}{
	                $current_c \leftarrow 1$\;
	            }{
	                $current_c \leftarrow C(pred_{AT})$\;
	            }
	        }
	        Create new state $s$\;
	        $S_+ \leftarrow S_+ \cup~\{s\}$\;
	        $T \leftarrow T \cup (pred_{AT}, activity(e), s)$\;
	        $A \leftarrow A \cup \{activity(e)\}$\;
	        $C \leftarrow C \cup (s, current_c)$\;
	        $current_{AT} \leftarrow s$\;
	    }
	    $seq(current_{AT}) \leftarrow seq(current_{AT}) \cup e$\;
	    $last_{AT}(case(e)) \leftarrow current_{AT}$\;
	}
}
\end{algorithm}

Lines 1 and 2 initialize variables. From Line 3 on, we iterate over the complete set of events of the plain event log. Lines 4--8 are concerned with identifying if the current event continues an already stored case or starts a new one at the root. Lines 10--12 map the resulting state to $current_{AT}$ for the case that a transition from a preceding state using the activity type of the current event already exists. If not, Lines 13---20 and Line 26 show how partitioning is performed. As a new state is created, it is immediately assigned to a partition and this assignment does not change. As Lines 14, 17, and 19 show, there are three possible cases of how a state can be assigned to a new partition:
\begin{enumerate}
    \item The preceding state $pred_{AT}$ already has some outgoing transitions. However, the $activity(e)$ is not among them. This means that a new path towards the next activity is introduced in $pred_{AT}$. This new path and the new states in this path are made a new partition and the partition count is incremented. 
    \item The preceding state is $root$ and it has no outgoing transitions, which means $e$ is the first event in $L_{plain}$. In this case, the new state defines the first partition in the extended prefix automaton. Note that this case does not differ from the previous one conceptually, but requires a slightly different implementation.
    \item The preceding state has no outgoing transitions but is not a root state. Since the $pred_{AT}$ has no outgoing transitions, the new activity does not add any path and thus the new state should belong to the same partition as its predecessor.
\end{enumerate}

We assume that all look-up functions in this algorithm can be implemented with computational complexity in $\mathcal{O}(k)$ with a constant $k$. Then, iterating over the set of events $E$ drives the complexity of this calculation. The complexity of calculating the extended prefix automaton is accordingly $\mathcal{O}(E)$.

\subsection{Graph Entropy of Extended Prefix Automata}\label{3.2}
Entropy is an appropriate concept for defining complexity measures due to its properties of monotonicity~\cite{polyvyanyy2020monotone}. Some applications of entropy have been developed by Polyvyanyy et al. for defining precision and recall measures for conformance checking at the language level, in which eigenvalues of process models and event logs are calculated iteratively with polynomial complexity~\cite{polyvyanyy2020monotone,polyvyanyy2020entropia,8786053}. Here, we highlight the opportunity to use entropy as an underlying concept for calculating the process complexity of an event log based on extended prefix automata in linear time.

More specifically, we define the four entropy measures: \textit{variant entropy} and \textit{sequence entropy}, as well as their corresponding normalized versions.
We build on the measures proposed by Dehmer et al.~\cite{dehmer2011history} who define graph entropy based on a partitioning of the graph into $X_i$ partitions.
\begin{equation}
    \label{eq:entropy}
    |X|\cdot log(|X|) - \sum\limits_{i=1}^{k} |X_i|\cdot log(|X_i|)
\end{equation}

We calculate \emph{variant entropy} based on the extended prefix automaton by only considering its structure and not the number of events associated with each state.
We apply the formula with $X = S$ (this is $S_+$ without the \emph{root}). Note that we do not consider the set of transitions $T$, because every state has exactly one incoming transition. We obtain:
\begin{equation}
\label{eq:graph_ent}
E_v = |S|\cdot log (|S|) - \sum\limits_{i=1}^{max(C)} |\{s \in S~|~C(s) = i\}|\cdot log(|\{s \in S~|~C(s) = i\}|)
\end{equation}

The measure of \textit{variant entropy} measures the complexity of an event log based the structure of the extended prefix automaton. It is possible to obtain the same value for two automata with a different number of states. 
For this reason, we introduce a \textit{normalized variant entropy} with a range of $[0,1]$. It is calculated as follows:
\begin{equation}
    \label{eq:norm_graph_ent}
    \bar{E_v} = \frac{E_v}{(|S|)\cdot log(|S|)}
\end{equation}

Both versions of the \textit{variant entropy} measure are based on the number of \textit{states} in a partition in~(\ref{eq:graph_ent}). This is an abstraction. Each state $s \in S$ is associated with a non-empty set of events, whose cases share the same prefix $seq(s) \neq \emptyset$. Each event $e \in L$ belongs to one and only one such set. This implies that every event can also be assigned to one and only one partition in the extended prefix automaton. In this way, we obtain a measure that reflects the frequencies of events and corresponding prefixes.
Thus, we calculate \textit{sequence entropy} based on the number of \textit{events} in a partition by extending~(\ref{eq:graph_ent}).
For the sake of readability, we define:
\begin{equation}
seq(S) = \bigcup\limits_{s \in S} seq(s)
\end{equation}
and
\begin{equation}
seq_i(S) = \bigcup\limits_{s \in S | C(s) = i} seq(s)
\end{equation}
in the following way:
\begin{equation}
    \label{eq:sequence_ent}
    E_s = |seq(S)|\cdot log(|seq(S)|) -
    \sum\limits_{i=1}^{max(C)} |seq_i(S) |\cdot log(|seq_i(S) |)
\end{equation}

Same as \textit{variant entropy}, the absolute \textit{sequence entropy} measure depends on the number of states in the extended prefix automaton, but additionally also on the number of events associated with each state. 
The same idea of normalization can be applied to \textit{sequence entropy} resulting in the \textit{normalized sequence entropy}:
\begin{equation}
    \label{eq:norm_sequence_ent}
    \bar{E_s} = \frac{E_s}{|seq(S)|\cdot log (|seq(S)|)}
\end{equation}

\begin{lemma}
The sequence entropy measure is monotonous with respect to an increasing number of events.
\end{lemma}
This property makes sequence entropy particularly suitable for measuring process complexity based on event logs.

\begin{proof}
In order to prove monotonicity of the sequence entropy measure, we have to show that adding one event increases this measure. To this end, we have to show that the following equation holds for $|seq(S_2)| = |seq(S_1)| +1$.

\begin{eqnarray}
    |seq(S_2)|\cdot log(|seq(S_2)|) -
    \sum\limits_{i=1}^{max(C_2)} |seq_i(S_2) |\cdot log(|seq_i(S_2) |) > \\
    |seq(S_1)|\cdot log(|seq(S_1)|) -
    \sum\limits_{i=1}^{max(C_1)} |seq_i(S_1) |\cdot log(|seq_i(S_1) |)
\end{eqnarray}

We observe a corner case. If $max(C_2)=max(C_1)=1$, then each summation equals the preceding term, such that both the left-hand and the right-hand side of the equation yield zero.

We rearrange the equation by bringing the sums onto the right-hand side.
\begin{eqnarray}
    |seq(S_2)|\cdot log(|seq(S_2)|) - |seq(S_1)|\cdot log(|seq(S_1)|) > \\
    \sum\limits_{i=1}^{max(C_2)} |seq_i(S_2) |\cdot log(|seq_i(S_2) |) - \\
    \sum\limits_{i=1}^{max(C_1)} |seq_i(S_1) |\cdot log(|seq_i(S_1) |)
\end{eqnarray}

Now we can distinguish two cases. If $max(C_2)>max(C_1)$, then there must be one new partition that includes only one event. As a result, the right-hand side becomes $1\cdot log(1)=0$, such that the formula holds true because $|seq(S_2)|\cdot log(|seq(S_2)|)$ is larger than $|seq(S_1)|\cdot log(|seq(S_1)|)$ due to each of its factors being larger, respectively.

Let us consider the alternative case that $max(C_2)=max(C_1)$. In this case, there exists an index $x$ where there is a difference between $seq_i(S_2)$ and $seq_i(S_1)$. We write
\begin{eqnarray}
    |seq(S_2)|\cdot log(|seq(S_2)|) - |seq(S_1)|\cdot log(|seq(S_1)|) > \\
    |seq(S_{x2})|\cdot log(|seq(S_{x2})|) - |seq(S_{x1})|\cdot log(|seq(S_{x1})|)
\end{eqnarray}

Furthermore, we can assume that there is a natural number $m < S_1$, such that we can write
\begin{eqnarray}
    |seq(S_2)|\cdot log(|seq(S_2)|) - |seq(S_1)|\cdot log(|seq(S_1)|) > \\
    (|seq(S_2)|-m)\cdot log(|seq(S_2)|-m) - (|seq(S_1)|-m)\cdot log(|seq(S_1)|-m)
\end{eqnarray}
We bring all terms to the left-hand side, yielding
\begin{eqnarray}
    |seq(S_2)|\cdot log(|seq(S_2)|) \\
    - (|seq(S_2)|-m)\cdot log(|seq(S_2)|-m) \\
    - |seq(S_1)|\cdot log(|seq(S_1)|)  \\
    + (|seq(S_1)|-m)\cdot log(|seq(S_1)|-m) \\
    > 0
\end{eqnarray}
Now, let us consider that we can write $S_2 = S_1\cdot f=S_1 + 1$, such that $f=\frac{S_1+1}{S_1}$. Then, we obtain
\begin{eqnarray}
    f\cdot|seq(S_1)|\cdot log(f\cdot|seq(S_1)|) \\
    - (f\cdot|seq(S_1)|-m)\cdot log(f\cdot|seq(S_1)|-m) \\
    - |seq(S_1)|\cdot log(|seq(S_1)|)  \\
    + (|seq(S_1)|-m)\cdot log(|seq(S_1)|-m) \\
    > 0
\end{eqnarray}
We pull $f$ out of the logarithms and reorder the terms to obtain
\begin{eqnarray}
    f\cdot|seq(S_1)|\cdot log(f)|)  \\
    - (f\cdot|seq(S_1)|-m)\cdot log(f) \\
    + f\cdot|seq(S_1)|\cdot log(|seq(S_1)|) \\
    - |seq(S_1)|\cdot log(|seq(S_1)|) \\
    - (f\cdot|seq(S_1)|-m)\cdot log(|seq(S_1)|-m) \\
    + (|seq(S_1)|-m)\cdot log(|seq(S_1)|-m) \\
    > 0
\end{eqnarray}
These can be pulled together to obtain
\begin{eqnarray}
    f\cdot log(f)\cdot ( |seq(S_1)| - |seq(S_1)|+m) \\
    + (f-1)\cdot|seq(S_1)|\cdot log(|seq(S_1)|) \\
    - (f-1)\cdot (|seq(S_1)|-m)\cdot log(|seq(S_1)|-m) \\
    > 0
\end{eqnarray}
This can be simplified and rewritten to
\begin{eqnarray}
    f\cdot log(f)\cdot m \\
    + (f-1)\cdot|seq(S_1)|\cdot log(|seq(S_1)|) \\
    > (f-1)\cdot (|seq(S_1)|-m)\cdot log(|seq(S_1)|-m)
\end{eqnarray}
We can replace the product on the right-hand side of the inequality with a term that is larger by replacing factors that are greater one with larger factors. In turn, we replace $|seq(S_1)|-m$ with $|seq(S_1)|$, such that the terms from lines (39) and (40) become equal. We then obtain
\begin{eqnarray}
    f\cdot log(f)\cdot m > 0
\end{eqnarray}
This equation holds true because the three factors in line (41) are greater than zero due to properties of $f$ and $m$. \qed
\end{proof}


Let us revisit \textbf{Observations 1--3} and focus on the sequence entropy measure.
\textbf{Observation 1} states that we would want to distinguish two event logs $L_1$ and $L_2$ that have the same amount of events $n=|E|$ and sequences $m=|\{\sigma~|~\ \sigma\in L\}|$. If all sequences of $L_1$ are identical, we obtain one partition. Based on the corresponding extended prefix automaton, we obtain $E_s = |seq(S)|\cdot log(|seq(S)|) -
    \sum\limits_{i=1}^{1} |seq_i(S) |\cdot log(|seq_i(S) |) = 0$.
If all sequences are fully different in $L_2$ (and let us assume them to be of equal length $l$), then we obtain
 $E_s = m\cdot l\cdot log(m\cdot l|) -
    \sum\limits_{i=1}^{m} l \cdot log(l) =
    m\cdot l\cdot log(|m\cdot l|) -
    m\cdot l \cdot log(l)=
    m\cdot l\cdot log(m) > 0$.
This means that these cases can be distinguished.

\textbf{Observation 2} states event logs $L_1$ and $L_2$ can have the same number of variants, but different size. Let us assume that $L_1$ is a duplication of $L_2$ such that $|L_1| = 2\cdot |L_2|$. Then, the amount of variants remains the same: $max(C_1)=max(C_2)$. For evaluating the impact on the entropy measure, we calculate the difference $E_s(L_1)-E_s(L_2)$. We assume $S$ refers to $L_2$, which yields the following equation: $2\cdot |S|\cdot log(2\cdot |S|) - \sum\limits_{i=1}^{max(C)} 2\cdot|seq_i(S) |\cdot log(2\cdot|seq_i(S) |) - |S|\cdot log(\cdot |S|) + \sum\limits_{i=1}^{max(C)} |seq_i(S) |\cdot log(\cdot |seq_i(S) |)$, which is equal to $2\cdot |S| \cdot log(2) + |S|\cdot log(|S|) - \sum\limits_{i=1}^{max(C)} 2\cdot |seq_i(S)| \cdot log(2) + |seq_i(S)|\cdot log(|seq_i(S)|)$. This is the same as $E_s + 2\cdot |S| \cdot log(2) - \sum\limits_{i=1}^{max(C)} 2\cdot |seq_i(S)| \cdot log(2)$ and greater than zero if $E_s$ is greater than one.

\textbf{Observation 3} states that two logs $L_1$ and $L_2$ with the same edit distance between cases can differ in size and in their number of variants. Regarding size, we already demonstrated that duplicating the event log increases entropy if there is more than one partition. If we assume a case of a constant number of states and events in the extended prefix automaton, then it suffices to show that two times the amount of events in one partition ($2\cdot |seq_i(S)|$) yields a higher entropy than two partitions with $|seq_i(S)|$ events. For these two cases, we have summands in the summation as follows: $(2\cdot |seq_i(S)|)\cdot log(2\cdot |seq_i(S)|)$ for less and larger partitions and $2\cdot (|seq_i(S)|\cdot log(|seq_i(S)|))$ for double the amount of partitions with half the number of events. It is easy to see that

\begin{equation}
    (2\cdot |seq_i(S)|)\cdot log(2\cdot |seq_i(S)|) > 2\cdot (|seq_i(S)|\cdot log(|seq_i(S)|))
\end{equation}
because
\begin{equation}
    2\cdot |seq_i(S)|\cdot 2 + 2\cdot |seq_i(S)|\cdot log(|seq_i(S)|) > 2\cdot (|seq_i(S)|\cdot log(|seq_i(S)|))
\end{equation}
yields
\begin{equation}
    2\cdot |seq_i(S)|\cdot 2 > 0
\end{equation}

To summarize, we observe that our sequence entropy yields a process complexity measure for event logs that is monotonously growing with events being added no matter where in the event log. In comparison to the other measures presented in Section \Ref{sec:back}, we observe that the critical Observations 1--3 are well addressed by our entropy-based measure.

\section{Evaluation}~\label{sec:eval}
We have implemented our graph entropy-based process complexity measures as a Python application,\footnote{Sources available at \url{https://github.com/MaxVidgof/process-complexity}} which also computes all log complexity measures discussed in Section~\ref{sec:back}. 
The application receives as input an event log (either in CSV or XES format) and calculates all complexity measures or a user-selected subset of them. 
Henceforth, for simplicity, we will refer to these measures as \emph{log complexity} measures to distinguish them from the complexity that is associated with process models.
We focus our evaluation and the analysis of the different complexity measures on the following two research questions:

\begin{itemize}
    \item[\textbf{RQ1.}] How does log complexity affect the quality of automatically discovered process models?
    \item[\textbf{RQ2.}] What log complexity measures could be used as a proxy to predict the quality of automatically discovered process models?
\end{itemize}

In the following, we describe in detail the analysis that we conducted towards answering the two research questions.

\subsection{Dataset and Setup}
For our experiments, we first selected the collection of 24 event logs that Augusto et al. used for their review~\cite{augusto2018automated}. We extended this collection with eight event logs of the Business Process Intelligence Challenges 2019~\cite{vandongen_bpi19} (three out of eight) and 2020~\cite{vandongen_bpi20} (five out of eight). Hence, the collection of event logs we use for our experiments includes a total of 32 event logs (20 of which are publicly available and 12 private). To the best of our knowledge, this is the largest collection of real-world event logs used in a process mining study so far, representing an increase of $33.3\%$ from the former largest collection~\cite{augusto2018automated}.
The public subset of the collection of event logs previously used by Augusto et al.~\cite{augusto2018automated} is available for download from the 4TU Research Data Centre~\cite{augusto2019dataset}. 
The event logs we used in our experiments record the executions of business processes from a variety of domains, including healthcare, finance, government, and IT service management. They demonstrate an heterogeneous degree of complexity across the different complexity measures. 
Tables~\ref{tab:logs-complexity1}--\ref{tab:logs-complexity3} show the complexity of each event log, reporting our novel graph entropy-based complexity measures (Table~\ref{tab:logs-complexity1}) as well as the complexity measures from prior research (Tables~\ref{tab:logs-complexity2} and~\ref{tab:logs-complexity3}). The labels of the state-of-the-art log complexity measures are reported in Table~\ref{tab:measures} in Section~\ref{sec:back}, while the labels of our measures are reported in Table~\ref{tab:ourmeasures-acronyms}.
\begin{wraptable}[5]{r}{0.50\textwidth}
\vspace{-3mm}
{\scriptsize{
\centering
    \begin{tabular}{l|l}
    \hline
          \textbf{Process Complexity Measure}
          & \textbf{Label}\\\hline

    \emph{Variant Entropy} & var-e\\

    \emph{Sequence Entropy} & seq-e\\

    \emph{Normalised Variant Entropy} & nvar-e\\

    \emph{Normalised Sequence Entropy} & nseq-e\\\hline

        \end{tabular}%
\vspace{-2mm}
  \caption{Process complexity measure labels}
  \label{tab:ourmeasures-acronyms}%
}}
\end{wraptable}%
The values\footnote{A dash '-' represents an exception during the calculation of the measure, e.g. due to out-of-memory exception or a timeout (which we set at 12 hours).} shown in Tables~\ref{tab:logs-complexity1}--\ref{tab:logs-complexity3} highlight the heterogeneous nature of the event logs in the collection, with the vast majority of the log complexity measures varying over a large range of values (e.g. \emph{DT(\%)} covers the range 0.01\% to 97.5\%).

\begingroup
\begin{table}[tbp]
\makebox[\textwidth][c]{
\centering
{\scriptsize{
    \begin{tabular}{l|r|r|r|r|r|r}

    \textbf{log} &
    \textbf{var-e} &
    \textbf{seq-e} &
    \textbf{nvar-e} &
    \textbf{nseq-e} &
    \textbf{LOD} &
    \textbf{granularity} \\\hline

\textbf{BPIC12} & 474928.8 & 1384057.4 & 0.708 & 0.423 & 8.2   & 0.26 \\
    \textbf{BPIC13cp} & 3502.3 & 18231.6 & 0.705 & 0.311 & 2.4   & 2502013.36 \\
    \textbf{BPIC13inc} & 88677.4 & 294092.5 & 0.718 & 0.405 & 2.6   & 1540.64 \\
    \textbf{BPIC14} & 1114386.4 & 2490701.5 & 0.772 & 0.526 & 5.5   & 139.90 \\
    \textbf{BPIC15f1} & 20385.6 & 90549.4 & 0.652 & 0.419 & 24.0  & 95.79 \\
    \textbf{BPIC15f2} & 50042.2 & 120443.3 & 0.640 & 0.483 & \textbf{36.1} & 761.23 \\
    \textbf{BPIC15f3} & 78199.7 & 231332.5 & 0.690 & 0.494 & 31.8  & 63.11 \\
    \textbf{BPIC15f4} & 34433.4 & 131444.0 & 0.664 & 0.434 & 34.1  & 0.01 \\
    \textbf{BPIC15f5} & 39903.1 & 134870.6 & 0.661 & 0.436 & 30.8  & \textbf{0.00} \\
    \textbf{BPIC17} & 505341.9 & 3443057.9 & 0.777 & 0.358 & 14.4  & \textbf{0.00} \\
    \textbf{BPIC19c1} & 1152607.5 & 1776714.2 & 0.598 & 0.499 & 3.7   & 434462.49 \\
    \textbf{BPIC19c2} & 56603.5 & 2475331.5 & 0.799 & 0.183 & 4.3   & 347055.36 \\
    \textbf{BPIC19c3} & 1194.1 & 11332.1 & 0.633 & 0.264 & 2.9   & 13745.14 \\
    \textbf{BPIC20a} & 1649.7 & 101733.0 & 0.696 & 0.165 & 5.2   & 11984.67 \\
    \textbf{BPIC20b} & 30722.0 & 273919.7 & 0.758 & 0.339 & 10.7  & 1746.48 \\
    \textbf{BPIC20c} & 96618.4 & 413564.3 & 0.734 & 0.420 & 10.8  & 2573.42 \\
    \textbf{BPIC20d} & 5488.8 & 56758.6 & 0.724 & 0.317 & 8.5   & 236.00 \\
    \textbf{BPIC20e} & 1322.8 & 73131.6 & 0.704 & 0.189 & 5.2   & 3163.74 \\
    \textbf{RTFMP} & 1892.3 & 831983.5 & 0.769 & 0.112 & 3.7   & 2174191.14 \\
    \textbf{SEPSIS} & 40624.5 & 76528.7 & 0.696 & 0.522 & 9.1   & 5.56 \\
    \textbf{PRT1} & 36311.9 & 236957.6 & 0.770 & 0.280 & 4.5   & 7689.03 \\
    \textbf{PRT2} & 255222.9 & 302150.0 & 0.638 & \textbf{0.608} & 7.9   & 0.01 \\
    \textbf{PRT3} & 31408.0 & 64170.9 & 0.656 & 0.491 & 8.4   & \textbf{0.00} \\
    \textbf{PRT4} & 305234.0 & 1035205.9 & 0.818 & 0.518 & 7.1   & 738.05 \\
    \textbf{PRT5} & \textbf{4.5} & \textbf{2050.0} & \textbf{0.270} & \textbf{0.055} & 6.0   & 590.90 \\
    \textbf{PRT6} & 4079.4 & 19076.6 & 0.750 & 0.365 & 7.5   & 665.25 \\
    \textbf{PRT7} & 3013.1 & 54130.7 & 0.745 & 0.341 & 8.2   & 0.00 \\
    \textbf{PRT8} & 38225.6 & 44186.7 & 0.547 & 0.534 & 6.1   & 474.93 \\
    \textbf{PRT9} & 21210.3 & 3628862.6 & 0.734 & 0.139 & 2.2   & 154278.43 \\
    \textbf{PRT10} & 949.6 & 214831.2 & 0.805 & 0.242 & \textbf{1.6} & \textbf{17685299.33} \\
    \textbf{PRT11} & \textbf{1207005.2} & \textbf{8588051.4} & 0.651 & 0.281 & 9.0   & \textbf{0.00} \\
    \textbf{PRT12} & 54394.4 & 541576.6 & \textbf{0.852} & 0.276 & 4.3   & 132703.06 \\
\hline
    \end{tabular}%
    }}}
  \caption{Complexity of the evaluation dataset - Part 1}
  \label{tab:logs-complexity1}%
\end{table}%
\endgroup

\begingroup
\begin{table}[tbp]
\makebox[\textwidth][c]{
\centering
{\scriptsize{
    \begin{tabular}{l|r|r|r|r|r|r|r}

    \textbf{log} &
    \textbf{structure} &
    \textbf{affinity} &
    \textbf{dev-random} &
    \textbf{avg-dist.} &
    \textbf{LZ} &
    \textbf{t-comp} &
    \textbf{magnitude}\\\hline

    \textbf{BPIC12} & 0.783 & 0.258 & 0.755 & 22.1  & 41378 & 173248 & 262200 \\
    \textbf{BPIC13cp} & 0.375 & 0.560 & 0.545 & 2.7   & 912   & 1715  & 6660 \\
    \textbf{BPIC13inc} & 0.313 & 0.644 & 0.577 & 6.7   & 6527  & 28280 & 65533 \\
    \textbf{BPIC14} & 0.580 & \textbf{0.142} & 0.842 & 8.3   & 62106 & 224591 & 369485 \\
    \textbf{BPIC15f1} & 0.975 & 0.229 & 0.880 & 20.6  & 3790  & 10367 & 21656 \\
    \textbf{BPIC15f2} & 0.971 & 0.208 & \textbf{0.903} & 29.8  & 4496  & 17820 & 24678 \\
    \textbf{BPIC15f3} & 0.951 & 0.216 & 0.895 & 23.3  & 7413  & 28553 & 43786 \\
    \textbf{BPIC15f4} & 0.964 & 0.243 & 0.894 & 24.7  & 4798  & 16502 & 29403 \\
    \textbf{BPIC15f5} & 0.972 & 0.224 & 0.894 & 26.6  & 4731  & 18789 & 30030 \\
    \textbf{BPIC17} & 0.904 & 0.689 & 0.734 & 13.8  & 84060 & \textbf{283587} & 714198 \\
    \textbf{BPIC19c1} & 0.240 & 0.411 & 0.481 & 23.8  & 18367 & 221868 & 283407 \\
    \textbf{BPIC19c2} & 0.344 & 0.394 & 0.647 & 2.3   & 53117 & 21295 & 979942 \\
    \textbf{BPIC19c3} & 0.375 & 0.466 & 0.580 & 3.5   & \textbf{593} & 625   & 5038 \\
    \textbf{BPIC20a} & 0.865 & 0.541 & 0.617 & 1.3   & 7947  & 734   & 56437 \\
    \textbf{BPIC20b} & 0.830 & 0.406 & 0.770 & 5.1   & 13401 & 9582  & 72151 \\
    \textbf{BPIC20c} & 0.787 & 0.303 & 0.800 & 8.2   & 16519 & 25074 & 86581 \\
    \textbf{BPIC20d} & 0.806 & 0.395 & 0.740 & 3.4   & 3932  & 1937  & 18246 \\
    \textbf{BPIC20e} & 0.878 & 0.515 & 0.622 & 1.4   & 5596  & 618   & 36796 \\
    \textbf{RTFMP} & 0.421 & 0.367 & 0.602 & 8.6   & 38212 & 1166  & 561470 \\
    \textbf{SEPSIS} & 0.551 & 0.256 & 0.787 & 11.8  & 3115  & 12956 & 15214 \\
    \textbf{PRT1} & 0.543 & 0.566 & 0.647 & 2.4   & 12388 & 13419 & 75353 \\
    \textbf{PRT2} & \textbf{0.025} & 0.294 & 0.821 & 36.5  & 8826  & 45864 & 46282 \\
    \textbf{PRT3} & 0.742 & 0.246 & 0.763 & 3.2   & 1541  & 13650 & 13720 \\
    \textbf{PRT4} & 0.694 & 0.253 & 0.795 & 5.2   & 32584 & 68432 & 166282 \\
    \textbf{PRT5} & 0.806 & \textbf{0.714} & 0.636 & \textbf{0.7} & 904   & \textbf{12} & \textbf{4434} \\
    \textbf{PRT6} & 0.728 & 0.399 & 0.747 & 2.9   & 1471  & 1691  & 6011 \\
    \textbf{PRT7} & 0.775 & 0.323 & 0.762 & 2.5   & 4183  & 1168  & 16353 \\
    \textbf{PRT8} & 0.859 & 0.155 & 0.831 & \textbf{51.7} & 2846  & 8950  & 9086 \\
    \textbf{PRT9} & 0.063 & 0.482 & \textbf{0.437} & 2.0   & 141857 & 7278  & 1808706 \\
    \textbf{PRT10} & 0.798 & 0.142 & 0.626 & 1.6   & 6976  & 356   & 78864 \\
    \textbf{PRT11} & \textbf{0.982} & -     & 0.839 & 18.2  & \textbf{165919} & 278322 & \textbf{2099835} \\
    \textbf{PRT12} & 0.355 & 0.236 & 0.730 & 2.3   & 28888 & 17397 & 163224 \\\hline
    \end{tabular}%
    }}}
  \caption{Complexity of the evaluation dataset - Part 2}
  \label{tab:logs-complexity2}%
\end{table}%
\endgroup

\begingroup
\begin{table}[tbp]
\makebox[\textwidth][c]{
\centering
{\scriptsize{
    \begin{tabular}{l|r|r|r|r|r|r|r}

    \textbf{log} &
    \textbf{support} &
    \textbf{variety} &
    \textbf{DT (\%)} &
    \textbf{DT (\#)} &
    \textbf{TL-min} &
    \textbf{TL-avg} &
    \textbf{TL-max} \\\hline

    \textbf{BPIC12} & 13087 & 24    & 33.4  & 4371  & 3     & 20    & 175 \\
    \textbf{BPIC13cp} & 1487  & \textbf{4} & 12.3  & 183   & \textbf{1} & 4     & 35 \\
    \textbf{BPIC13inc} & 7554  & \textbf{4} & 20.0  & 1511  & \textbf{1} & 9     & 123 \\
    \textbf{BPIC14} & 41353 & 9     & 36.1  & \textbf{14928} & 3     & 9     & 167 \\
    \textbf{BPIC15f1} & 902   & 70    & 32.7  & 295   & 5     & 24    & 50 \\
    \textbf{BPIC15f2} & 681   & 82    & 61.7  & 420   & 4     & 36    & 63 \\
    \textbf{BPIC15f3} & 1369  & 62    & 60.3  & 826   & 4     & 32    & 54 \\
    \textbf{BPIC15f4} & 860   & 65    & 52.4  & 451   & 5     & 34    & 54 \\
    \textbf{BPIC15f5} & 975   & 74    & 45.7  & 446   & 4     & 31    & 61 \\
    \textbf{BPIC17} & 21861 & 18    & 40.1  & 8766  & 11    & 33    & 113 \\
    \textbf{BPIC19c1} & 15129 & 5     & 20.9  & 3159  & \textbf{1} & 19    & 794 \\
    \textbf{BPIC19c2} & 220810 & 8     & 1.2   & 2706  & \textbf{1} & 4     & 179 \\
    \textbf{BPIC19c3} & 1027  & \textbf{4} & 10.1  & 104   & 2     & 5     & 19 \\
    \textbf{BPIC20a} & 10500 & 17    & 0.9   & 99    & \textbf{1} & 5     & 24 \\
    \textbf{BPIC20b} & 6449  & 34    & 11.7  & 753   & 3     & 11    & 7 \\
    \textbf{BPIC20c} & 7065  & 51    & 20.9  & 1478  & 3     & 12    & 90 \\
    \textbf{BPIC20d} & 2099  & 29    & 9.6   & 202   & \textbf{1} & 9     & 21 \\
    \textbf{BPIC20e} & 6886  & 19    & 1.3   & 89    & \textbf{1} & 5     & 20 \\
    \textbf{RTFMP} & 150370 & 11    & 0.2   & 301   & 2     & 4     & 20 \\
    \textbf{SEPSIS} & 1050  & 16    & 80.6  & 846   & 3     & 14    & 185 \\
    \textbf{PRT1} & 12720 & 9     & 8.1   & 1030  & 2     & 5     & 64 \\
    \textbf{PRT2} & 1182  & 9     & 97.5  & 1152  & \textbf{12} & 39    & 276 \\
    \textbf{PRT3} & 1600  & 15    & 19.9  & 318   & 6     & 8     & 9 \\
    \textbf{PRT4} & 20000 & 11    & 29.7  & 5940  & 6     & 8     & 36 \\
    \textbf{PRT5} & 739   & 6     & 0.10  & \textbf{1} & 6     & 6     & \textbf{6} \\
    \textbf{PRT6} & 744   & 9     & 22.4  & 167   & 7     & 8     & 21 \\
    \textbf{PRT7} & 2000  & 13    & 6.4   & 128   & 8     & 8     & 11 \\
    \textbf{PRT8} & \textbf{225} & 55    & \textbf{99.9} & 225   & 2     & \textbf{40} & 350 \\
    \textbf{PRT9} & \textbf{787657} & 8     & \textbf{0.01} & 79    & \textbf{1} & 2     & 58 \\
    \textbf{PRT10} & 43514 & 19    & \textbf{0.01} & 4     & \textbf{1} & \textbf{1} & 15 \\
    \textbf{PRT11} & 174842 & \textbf{310} & 3.0   & 5245  & 2     & 12    & \textbf{804} \\
    \textbf{PRT12} & 37345 & 20    & 7.5   & 2801  & \textbf{1} & 4     & 27 \\\hline
    \end{tabular}%
    }}}
  \caption{Complexity of the evaluation dataset - Part 3}
  \label{tab:logs-complexity3}%
\end{table}%
\endgroup
Given our interest in understanding how log complexity measures relate to the quality of automatically discovered process models, the second component of our evaluation dataset includes the process models automatically discovered from the 32 event logs using the three most reliable algorithms to date, according to~\cite{augusto2018automated}: the Evolutionary Tree Miner (ETM)~\cite{BuijsDV12}, the Inductive Miner - Infrequent Behavior Variant (IM)~\cite{leemans2014infrequent}, and the Split Miner (SM)~\cite{augusto2018split}.

Table~\ref{tab:models-quality} shows the quality of the process models automatically discovered by ETM, IM, and SM, covering both accuracy (i.e., fitness and precision) and complexity of the process models.
The corresponding measures are: \emph{fitness}, \emph{precision}, \emph{F-score} (of fitness and precision), \emph{size}, and \emph{control flow complexity} (CFC).
We recall that \emph{fitness} quantifies the amount of behavior contained in the event log that the process model is able to replay. The \emph{precision} measure quantifies the amount of behavior allowed by the process model that can be found in the event log.
The F-score of fitness and precision is the product of the two measurements divided by their sum and multiplied by two.
Over the past decade, several fitness and precision measures have been proposed~\cite{syring11790evaluating}, each of them suffering from different limitations (from approximation to low scalability). The results shown in Table~\ref{tab:models-quality} report the alignment-based fitness and precision measures proposed by Adriansyah et al.~\cite{adriansyah2011conformance, adriansyah2015measuring}. The choice of measures was guided by the goal to maintain consistency with the latest benchmark results~\cite{augusto2018automated}. The \emph{alignment-based fitness}~\cite{adriansyah2011conformance}
is calculated as one minus the normalized sum of the minimal alignment cost between each trace in the event log and the closest corresponding trace that can be replayed by the process model.
The \emph{alignment-based precision}~\cite{adriansyah2015measuring}, instead, builds a prefix-automaton of the event log and then replays the process model on top of it, assessing the number of times that the process behavior diverges from the behavior of the prefix-automation.
Finally, the \emph{size} of a process model (in BPMN format\footnote{\url{bpmn.org}}) is equal to the number of nodes composing the process model, while the \emph{CFC} of a (BPMN) process model calculates the amount of branching induced by its split gateways~\cite{DBLP:books/sp/Mendling2008}.

\begingroup
\begin{table}[tbp]
\makebox[\textwidth][c]{
\centering
{\scriptsize{
    \begin{tabular}{l|r|r|r|r|r|r|r|r|r|r|r|r|r|r|r}

          & \multicolumn{5}{c|}{ETM}
          & \multicolumn{5}{c|}{IM}
          & \multicolumn{5}{c}{SM} \\\cline{2-16}

          \textbf{Log}
          & \textbf{Fit.}
          & \textbf{Prec.}
          & \textbf{F-score}
          & \textbf{Size}
          & \textbf{CFC}
          & \textbf{Fit.}
          & \textbf{Prec.}
          & \textbf{F-score}
          & \textbf{Size}
          & \textbf{CFC}
          & \textbf{Fit.}
          & \textbf{Prec.}
          & \textbf{F-score}
          & \textbf{Size}
          & \textbf{CFC}  \\\hline

\textbf{BPIC12} & 0.33  & 0.98  & 0.49  & 69    & 10    & 0.98  & 0.50  & 0.66  & 59    & 37    & 0.96  & 0.81  & 0.88  & 51    & 41 \\
    \textbf{BPIC13cp} & 0.99  & 0.76  & 0.86  & 11    & 17    & \textbf{0.82} & \textbf{1.00} & 0.90  & \textbf{9} & 4     & 0.99  & 0.94  & 0.96  & 11    & 6 \\
    \textbf{BPIC13inc} & 0.84  & 0.80  & 0.82  & 28    & 24    & 0.92  & 0.56  & 0.70  & 13    & 7     & 0.98  & 0.92  & 0.95  & 12    & 8 \\
    \textbf{BPIC14} & 0.68  & 0.94  & 0.79  & 22    & 15    & 0.89  & 0.65  & 0.75  & 31    & 18    & 0.77  & 0.93  & 0.84  & 20    & 14 \\
    \textbf{BPIC15f1} & 0.57  & 0.89  & 0.69  & 73    & 21    & 0.97  & 0.57  & 0.71  & 164   & 108   & 0.90  & 0.90  & 0.90  & 111   & 45 \\
    \textbf{BPIC15f2} & 0.62  & 0.90  & 0.73  & 78    & 19    & 0.93  & 0.56  & 0.70  & 193   & 123   & 0.77  & 0.91  & 0.83  & 126   & 45 \\
    \textbf{BPIC15f3} & 0.66  & 0.88  & 0.75  & 78    & 26    & 0.95  & 0.55  & 0.70  & 159   & 108   & 0.79  & 0.93  & 0.85  & 94    & 33 \\
    \textbf{BPIC15f4} & 0.66  & 0.95  & 0.78  & 74    & 17    & 0.96  & 0.59  & 0.73  & 162   & 111   & \textbf{0.73} & 0.90  & 0.81  & 100   & 35 \\
    \textbf{BPIC15f5} & 0.58  & 0.89  & 0.70  & 82    & 26    & 0.94  & \textbf{0.18} & \textbf{0.30} & 134   & 95    & 0.79  & 0.94  & 0.86  & 106   & 34 \\
    \textbf{BPIC17} & 0.72  & \textbf{1.00} & 0.84  & 31    & 5     & 0.98  & 0.70  & 0.82  & 35    & 20    & 0.96  & 0.85  & 0.90  & 31    & 18 \\
    \textbf{BPIC19c1} & 0.30  & 1.00  & 0.46  & 33    & 4     & 0.97  & 0.37  & 0.54  & 19    & 11    & 0.85  & \textbf{0.61} & \textbf{0.71} & 17    & 9 \\
    \textbf{BPIC19c2} & 0.87  & 0.99  & 0.93  & 25    & 13    & 0.93  & 0.69  & 0.79  & 20    & 9     & 0.92  & 1.00  & 0.96  & 28    & 19 \\
    \textbf{BPIC19c3} & 0.78  & 0.77  & 0.78  & 21    & 9     & \textbf{1.00} & 0.77  & 0.87  & 13    & 7     & 0.89  & 0.98  & 0.93  & 12    & 6 \\
    \textbf{BPIC20a} & 0.88  & \textbf{0.43} & 0.58  & 23    & 10    & 0.95  & 0.86  & 0.90  & 34    & 19    & 0.97  & 0.78  & 0.87  & 33    & 16 \\
    \textbf{BPIC20b} & 0.77  & 0.55  & 0.64  & 55    & 25    & 0.88  & 0.55  & 0.68  & 88    & 49    & 0.96  & 0.84  & 0.90  & 70    & 43 \\
    \textbf{BPIC20c} & 0.67  & 0.92  & 0.77  & 68    & 32    & 0.74  & 0.32  & 0.44  & 100   & 62    & 0.97  & 0.78  & 0.86  & 146   & 113 \\
    \textbf{BPIC20d} & 0.80  & 0.98  & 0.88  & 44    & 19    & 0.91  & 0.44  & 0.60  & 66    & 37    & 0.95  & 0.86  & 0.90  & 58    & 31 \\
    \textbf{BPIC20e} & 0.89  & 0.88  & 0.88  & 35    & 16    & 0.92  & 0.87  & 0.90  & 38    & 21    & \textbf{1.00} & 0.87  & 0.93  & 37    & 19 \\
    \textbf{RTFMP} & 0.79  & 0.98  & 0.87  & 46    & 33    & 0.99  & 0.70  & 0.82  & 34    & 20    & \textbf{1.00} & 0.97  & 0.98  & 22    & 17 \\
    \textbf{SEPSIS} & 0.71  & 0.84  & 0.77  & 30    & 15    & 0.99  & 0.45  & 0.62  & 50    & 32    & 0.76  & 0.86  & 0.81  & 33    & 23 \\
    \textbf{PRT1} & 0.99  & 0.81  & 0.89  & 23    & 12    & 0.90  & 0.67  & 0.77  & 20    & 9     & 0.98  & 0.99  & 0.98  & 29    & 18 \\
    \textbf{PRT2} & 0.57  & 0.94  & 0.71  & \textbf{86} & 21    & -     & -     & -     & 45    & 33    & 0.81  & 0.74  & 0.77  & 40    & 30 \\
    \textbf{PRT3} & 0.98  & 0.86  & 0.92  & 51    & 37    & 0.98  & 0.68  & 0.80  & 37    & 20    & 0.83  & 0.91  & 0.87  & 32    & 17 \\
    \textbf{PRT4} & 0.84  & 0.85  & 0.84  & 64    & 28    & 0.93  & 0.75  & 0.83  & 27    & 13    & 0.87  & 0.99  & 0.93  & 34    & 21 \\
    \textbf{PRT5} & \textbf{1.00} & \textbf{1.00} & \textbf{1.00} & \textbf{10} & \textbf{1} & \textbf{1.00} & \textbf{1.00} & \textbf{1.00} & 10    & \textbf{1} & \textbf{1.00} & \textbf{1.00} & \textbf{1.00} & \textbf{10} & \textbf{1} \\
    \textbf{PRT6} & 0.98  & 0.80  & 0.88  & 41    & 16    & 0.99  & 0.82  & 0.90  & 23    & 10    & 0.94  & \textbf{1.00} & 0.97  & 16    & 5 \\
    \textbf{PRT7} & 0.90  & 0.81  & 0.85  & 60    & 29    & \textbf{1.00} & 0.73  & 0.84  & 29    & 13    & 0.91  & \textbf{1.00} & 0.95  & 30    & 11 \\
    \textbf{PRT8} & 0.35  & 0.88  & 0.50  & 75    & 12    & 0.98  & 0.33  & 0.49  & 111   & 92    & 0.97  & 0.67  & 0.79  & 406   & 488 \\
    \textbf{PRT9} & 0.75  & 0.49  & 0.59  & 27    & 13    & 0.90  & 0.61  & 0.73  & 28    & 16    & 0.92  & \textbf{1.00} & 0.96  & 30    & 20 \\
    \textbf{PRT10} & \textbf{1.00} & 0.63  & 0.77  & 61    & \textbf{45} & 0.96  & 0.79  & 0.87  & 41    & 29    & 0.97  & 0.97  & 0.97  & 79    & 68 \\
    \textbf{PRT11} & \textbf{0.10} & \textbf{1.00} & \textbf{0.18} & 21    & 3     & -     & -     & -     & \textbf{549} & \textbf{365} & -     & -     & -     & \textbf{712} & \textbf{609} \\
    \textbf{PRT12} & 0.63  & \textbf{1.00} & 0.77  & 21    & 8     & \textbf{1.00} & 0.77  & 0.87  & 32    & 25    & 0.96  & 0.99  & 0.97  & 97    & 84 \\
\hline

    \end{tabular}%
    }}}
  \caption{Quality measurements of the process models automatically discovered by ETM, IM, and SM}
  \label{tab:models-quality}%
\end{table}%
\endgroup

Starting from the log complexity measurements and the process model quality measurements, we calculated the Pearson correlation and the Kendall correlation~\cite{kendall1948rank} for each pair of corresponding measurement series (log complexity, process model quality) and assessed their statistical significance. While the Pearson correlation focuses on the likelihood of an existing linear relation between two measurement series (e.g., between the \emph{magnitude} of the event logs and the \emph{fitness} of the models discovered by IM), the Kendall correlation tells us whether two measurement series exhibit the same rank. The Kendall correlation is particularly useful when two measurement series do not exhibit the same trend (e.g., a linear or an exponential relation), yet they rank the objects of the measurements identically or similarly (to a certain degree, assessed via statistical significance). In our context, the Kendall correlation allows us to understand if a log complexity measure calculated for a set of logs ranks the logs identically or similarly to the rank yielded by a process model quality measure. 
 Furthermore, we conduct a regression analysis to study the potential to predict process model complexity based on log complexity.

We leverage the results of the correlation analysis to collect evidence that allowed us to answer RQ1. Then, we select a subset of the log complexity measures that correlated the most with the quality measures of automatically discovered process models, and we explore if and how they could be used to estimate apriori the quality of the automatically discovered process models (answering RQ2). Finally, we compare the time performance of our novel entropy-based log complexity measures against the existing log complexity measures.
All the experiments were run on an Intel Core i7-8565U@1.80GHz with 32GB RAM running Windows 10 Pro (64-bit) and Python 3.8.5 with no RAM limitation.

\subsection{Results of Correlation Analysis}

\begingroup
\begin{table}[tbp]
\makebox[\textwidth][c]{
\centering
{\scriptsize{
    \begin{tabular}{l|r|r|r|r|r|r|r|r|r|r|r|r|r|r|r}

          \textbf{Log Comp.}
          & \multicolumn{5}{c|}{\textbf{ETM}}
          & \multicolumn{5}{c|}{\textbf{IM}}
          & \multicolumn{5}{c}{\textbf{SM}} \\\cline{2-16}

          \textbf{Measure}
          & \textbf{Fit.}
          & \textbf{Prec.}
          & \textbf{F-score}
          & \textbf{Size}
          & \textbf{CFC}
          & \textbf{Fit.}
          & \textbf{Prec.}
          & \textbf{F-score}
          & \textbf{Size}
          & \textbf{CFC}
          & \textbf{Fit.}
          & \textbf{Prec.}
          & \textbf{F-score}
          & \textbf{Size}
          & \textbf{CFC}  \\\hline

    \textbf{var-e} & \textbf{-0.41} & \underline{0.34} & -0.16 & 0.16  & -0.08 & -0.17 & \textbf{-0.40} & \textbf{-0.40} & 0.07  & 0.12  & \underline{-0.29} & -0.23 & \textbf{-0.37} & 0.11  & 0.16 \\
    \textbf{seq-e} & -0.25 & 0.24  & -0.16 & 0.00  & -0.01 & \underline{-0.26} & -0.18 & -0.19 & -0.01 & 0.03  & -0.10 & -0.02 & -0.06 & 0.03  & 0.16 \\
    \textbf{nvar-e} & 0.24  & -0.03 & 0.19  & -0.16 & 0.13  & -0.18 & 0.22  & 0.20  & -0.17 & -0.15 & 0.10  & 0.21  & \textbf{0.38} & -0.07 & 0.05 \\
    \textbf{nseq-e} & \textbf{-0.39} & 0.10  & -0.25 & \textbf{0.37} & 0.13  & -0.04 & \textbf{-0.43} & \textbf{-0.43} & \underline{0.28} & \underline{0.30} & \textbf{-0.48} & \underline{-0.29} & \textbf{-0.61} & \underline{0.26} & 0.16 \\
    \textbf{LOD} & \underline{-0.31} & 0.20  & -0.09 & \textbf{0.46} & 0.15  & 0.00  & \underline{-0.32} & \underline{-0.29} & \textbf{0.59} & \textbf{0.56} & \textbf{-0.37} & -0.19 & \textbf{-0.43} & \textbf{0.45} & \underline{0.32} \\
    \textbf{granularity} & 0.19  & -0.17 & -0.02 & \underline{-0.32} & -0.10 & -0.18 & 0.22  & 0.19  & \underline{-0.30} & \underline{-0.27} & \underline{0.32} & 0.09  & \underline{0.32} & -0.19 & -0.07 \\
    \textbf{structure} & -0.17 & 0.08  & -0.12 & \textbf{0.46} & 0.17  & -0.01 & -0.10 & -0.09 & \textbf{0.60} & \textbf{0.54} & -0.06 & -0.23 & \underline{-0.27} & \textbf{0.49} & \underline{0.31} \\
    \textbf{affinity} & \textbf{0.34} & -0.10 & \underline{0.26} & \textbf{-0.52} & \underline{-0.29} & -0.04 & \underline{0.26} & 0.25  & \textbf{-0.46} & \textbf{-0.49} & \textbf{0.42} & 0.05  & \textbf{0.34} & \textbf{-0.49} & \textbf{-0.47} \\
    \textbf{dev-random} & \underline{-0.32} & 0.16  & -0.12 & \textbf{0.56} & 0.24  & -0.03 & \textbf{-0.34} & \underline{-0.32} & \textbf{0.60} & \textbf{0.59} & \textbf{-0.44} & -0.11 & \textbf{-0.44} & \textbf{0.50} & \textbf{0.39} \\
    \textbf{avg-dist.} & \textbf{-0.62} & 0.25  & \textbf{-0.37} & \textbf{0.50} & 0.08  & -0.03 & \textbf{-0.55} & \textbf{-0.55} & \textbf{0.40} & \textbf{0.43} & \textbf{-0.39} & \textbf{-0.36} & \textbf{-0.55} & \underline{0.31} & \underline{0.27} \\
    \textbf{LZ} & -0.17 & 0.20  & -0.14 & -0.03 & -0.03 & -0.25 & -0.11 & -0.12 & -0.02 & 0.02  & -0.02 & -0.04 & 0.03  & 0.01  & 0.14 \\
    \textbf{t-comp} & \textbf{-0.41} & \underline{0.33} & -0.13 & 0.17  & -0.03 & -0.17 & \textbf{-0.36} & \textbf{-0.36} & 0.10  & 0.15  & \underline{-0.32} & -0.18 & \textbf{-0.34} & 0.11  & 0.14 \\
    \textbf{magnitude} & -0.16 & 0.20  & -0.13 & -0.02 & 0.00  & -0.25 & -0.11 & -0.12 & -0.05 & -0.02 & -0.02 & -0.03 & 0.04  & 0.00  & 0.15 \\
    \textbf{support} & 0.07  & 0.04  & 0.02  & -0.24 & 0.00  & -0.15 & 0.08  & 0.06  & -0.24 & -0.21 & 0.16  & 0.08  & 0.24  & -0.17 & -0.01 \\
    \textbf{variety} & \textbf{-0.39} & 0.13  & \underline{-0.29} & \textbf{0.61} & 0.23  & -0.05 & \underline{-0.31} & \underline{-0.29} & \textbf{0.87} & \textbf{0.86} & -0.19 & \underline{-0.31} & \textbf{-0.37} & \textbf{0.81} & \textbf{0.63} \\
    \textbf{DT (\%)} & \textbf{-0.37} & 0.17  & -0.18 & \textbf{0.36} & 0.02  & 0.08  & \textbf{-0.33} & \underline{-0.33} & \underline{0.30} & \underline{0.32} & \textbf{-0.38} & -0.19 & \textbf{-0.43} & 0.24  & 0.18 \\
    \textbf{DT (\#)} & \underline{-0.30} & \textbf{0.35} & -0.05 & 0.06  & -0.06 & -0.13 & \underline{-0.30} & \underline{-0.30} & 0.03  & 0.07  & -0.24 & -0.18 & -0.25 & 0.03  & 0.10 \\
    \textbf{TL-min} & -0.08 & 0.10  & 0.12  & 0.27  & 0.16  & 0.27  & -0.01 & 0.02  & 0.22  & 0.18  & -0.26 & 0.10  & -0.09 & 0.08  & -0.03 \\
    \textbf{TL-avg} & \textbf{-0.50} & 0.23  & \underline{-0.30} & \textbf{0.48} & -0.01 & 0.00  & \textbf{-0.48} & \textbf{-0.45} & \textbf{0.47} & \textbf{0.46} & \textbf{-0.35} & \textbf{-0.40} & \textbf{-0.61} & 0.37  & \underline{0.26} \\
    \textbf{TL-max} & \textbf{-0.41} & 0.25  & \underline{-0.28} & 0.09  & -0.25 & -0.18 & \textbf{-0.41} & \textbf{-0.42} & 0.04  & 0.09  & -0.21 & \underline{-0.28} & \textbf{-0.39} & 0.11  & 0.16 \\\hline

    \end{tabular}%
    }}}
  \caption{Kendall correlation values. Bold and underlined values respectively highlight correlations at $p < 0.01$ and $p < 0.05$.}
  \label{tab:kendall-corr}%
\end{table}%
\endgroup
\begingroup
\begin{table}[tbp]
\makebox[\textwidth][c]{
\centering
{\scriptsize{
    \begin{tabular}{l|r|r|r|r|r|r|r|r|r|r|r|r|r|r|r}

          \textbf{Log Comp.}
          & \multicolumn{5}{c|}{\textbf{ETM}}
          & \multicolumn{5}{c|}{\textbf{IM}}
          & \multicolumn{5}{c}{\textbf{SM}} \\\cline{2-16}

          \textbf{Measure}
          & \textbf{Fit.}
          & \textbf{Prec.}
          & \textbf{F-score}
          & \textbf{Size}
          & \textbf{CFC}
          & \textbf{Fit.}
          & \textbf{Prec.}
          & \textbf{F-score}
          & \textbf{Size}
          & \textbf{CFC}
          & \textbf{Fit.}
          & \textbf{Prec.}
          & \textbf{F-score}
          & \textbf{Size}
          & \textbf{CFC}  \\\hline

     \textbf{var-e} & \textbf{-0.48} & 0.33  & -0.34 & -0.13 & -0.30 & -0.02 & -0.21 & -0.19 & -0.18 & -0.17 & -0.27 & \underline{-0.42} & \textbf{-0.50} & -0.17 & -0.10 \\
    \textbf{seq-e} & -0.21 & 0.07  & -0.17 & -0.27 & -0.31 & -0.07 & -0.05 & -0.01 & -0.27 & -0.28 & -0.02 & -0.01 & -0.02 & -0.23 & -0.14 \\
    \textbf{nvar-e} & 0.13  & -0.17 & 0.05  & 0.03  & \underline{0.36} & -0.24 & -0.01 & 0.04  & -0.14 & -0.16 & 0.04  & 0.23  & 0.21  & -0.19 & -0.18 \\
    \textbf{nseq-e} & \textbf{-0.52} & 0.26  & -0.32 & \textbf{0.53} & 0.17  & -0.06 & \textbf{-0.59} & \textbf{-0.56} & \underline{0.45} & \textbf{0.47} & \textbf{-0.64} & \underline{-0.41} & \textbf{-0.71} & \underline{0.37} & 0.28 \\
    \textbf{LOD} & -0.32 & 0.22  & -0.09 & \textbf{0.69} & 0.13  & 0.03  & \underline{-0.39} & \underline{-0.37} & \textbf{0.90} & \textbf{0.88} & \textbf{-0.65} & -0.02 & \underline{-0.42} & 0.29  & -0.01 \\
    \textbf{granularity} & 0.27  & -0.26 & 0.03  & 0.08  & \textbf{0.50} & 0.02  & 0.20  & 0.19  & -0.11 & -0.08 & 0.19  & 0.15  & 0.23  & 0.01  & 0.03 \\
    \textbf{structure} & -0.07 & 0.10  & 0.02  & \textbf{0.65} & 0.23  & 0.08  & -0.15 & -0.15 & \textbf{0.65} & \textbf{0.63} & -0.19 & -0.17 & -0.25 & \underline{0.39} & 0.20 \\
    \textbf{affinity} & \underline{0.43} & -0.15 & 0.30  & \textbf{-0.68} & \textbf{-0.46} & -0.08 & \underline{0.46} & \underline{0.41} & \textbf{-0.54} & \textbf{-0.57} & \textbf{0.56} & 0.09  & \underline{0.43} & \textbf{-0.49} & \underline{-0.37} \\
    \textbf{dev-random} & -0.30 & 0.33  & 0.00  & \textbf{0.71} & 0.24  & 0.04  & \underline{-0.41} & \underline{-0.37} & \textbf{0.76} & \textbf{0.75} & \textbf{-0.55} & -0.04 & \underline{-0.38} & \textbf{0.47} & 0.26 \\
    \textbf{avg-dist.} & \textbf{-0.79} & 0.31  & \textbf{-0.58} & \textbf{0.64} & -0.11 & 0.16  & \textbf{-0.65} & \textbf{-0.63} & \textbf{0.69} & \textbf{0.75} & \underline{-0.39} & \textbf{-0.56} & \textbf{-0.71} & \textbf{0.76} & \textbf{0.66} \\
    \textbf{LZ} & -0.13 & -0.10 & -0.17 & -0.24 & -0.22 & -0.11 & 0.00  & 0.04  & -0.24 & -0.24 & 0.06  & 0.14  & 0.14  & -0.19 & -0.11 \\
    \textbf{t-comp} & \underline{-0.46} & \underline{0.37} & -0.28 & -0.09 & -0.36 & 0.05  & -0.18 & -0.14 & -0.15 & -0.14 & -0.16 & \underline{-0.38} & \underline{-0.40} & -0.16 & -0.11 \\
    \textbf{magnitude} & -0.05 & -0.14 & -0.13 & -0.26 & -0.19 & -0.08 & 0.00  & 0.04  & -0.25 & -0.25 & 0.09  & 0.17  & 0.19  & -0.19 & -0.12 \\
    \textbf{support} & 0.04  & -0.36 & -0.16 & -0.21 & -0.07 & -0.12 & 0.02  & 0.04  & -0.19 & -0.18 & 0.08  & 0.26  & 0.25  & -0.13 & -0.08 \\
    \textbf{variety} & \textbf{-0.47} & 0.15  & -0.29 & \textbf{0.81} & 0.21  & -0.10 & \textbf{-0.56} & \textbf{-0.56} & \textbf{0.98} & \textbf{0.98} & \underline{-0.44} & -0.23 & \underline{-0.46} & \textbf{0.64} & 0.36 \\
    \textbf{DT (\%)} & \textbf{-0.48} & 0.23  & -0.27 & \textbf{0.49} & -0.06 & 0.22  & \underline{-0.45} & \underline{-0.41} & \textbf{0.58} & \textbf{0.63} & \textbf{-0.55} & -0.26 & \textbf{-0.56} & \textbf{0.60} & \textbf{0.52} \\
    \textbf{DT (\#)} & -0.22 & 0.32  & -0.02 & -0.18 & -0.22 & -0.09 & -0.02 & 0.00  & -0.19 & -0.19 & -0.22 & -0.06 & -0.18 & -0.17 & -0.10 \\
    \textbf{TL-min} & 0.10  & 0.25  & 0.29  & 0.26  & 0.03  & 0.31  & 0.09  & 0.14  & 0.14  & 0.11  & -0.22 & 0.18  & 0.01  & -0.07 & -0.14 \\
    \textbf{TL-avg} & \textbf{-0.66} & 0.34  & \underline{-0.40} & \textbf{0.66} & -0.12 & 0.15  & \textbf{-0.58} & \textbf{-0.55} & \textbf{0.78} & \textbf{0.81} & \textbf{-0.48} & \underline{-0.45} & \textbf{-0.66} & \textbf{0.62} & \underline{0.44} \\
    \textbf{TL-max} & \textbf{-0.65} & 0.29  & \textbf{-0.56} & -0.03 & \underline{-0.37} & 0.08  & \underline{-0.46} & \underline{-0.44} & -0.06 & -0.02 & -0.16 & \textbf{-0.71} & \textbf{-0.67} & 0.19  & 0.27 \\\hline

    \end{tabular}%
    }}}
  \caption{Pearson correlation values. Bold and underlined values respectively highlight correlations at $p < 0.01$ and $p < 0.05$.}
  \label{tab:pearson-corr}%
\end{table}%
\endgroup

Tables~\ref{tab:kendall-corr} and~\ref{tab:pearson-corr}, respectively, show the Kendall and Pearson correlations values of each pair of log complexity measure and process model quality measure (for a total of 300 pairs).
We highlight in bold and underline the correlation values that yield a p-value less than or equal to $0.01$ and a p-value less than or equal to $0.05$. Both thresholds are traditionally used to identify statistical significance.

We note that both the Pearson and Kendall correlations yield similar significance values, identifying 67 (Pearson) and 78 (Kendall) pairs of correlation measures with a p-value below $0.01$. The rate of overlap between the identified pairs is $75\%$ (54 pairs). As expected, we see that the Kendall correlation is looser than the Pearson correlation. If we consider pairs that correlate with a p-value at $0.05$, Kendall identifies 119 pairs while Pearson identifies 99 (17\% less), with an overlap of $83\%$ (90 pairs). Table~\ref{tab:pearken-corr} shows the overlap of statistical significance and the corresponding p-values. A pair of measures that correlate negatively is identified with the dash symbol between brackets. To reinforce the statistical significance, in the following, we refer only to the results reported in Table~\ref{tab:pearken-corr}. The correlation \emph{summary} reported in Table~\ref{tab:pearken-corr} can be used to analyze which log complexity measures affect the process model quality measures the most and vice-versa. This analysis allows us to answer RQ1.

\begingroup
\begin{table}[tbp]
\makebox[\textwidth][c]{
\centering
{\scriptsize{
    \begin{tabular}{l|l|l|l|l|l|l|l|l|l|l|l|l|l|l|l}

          \textbf{Log Comp.}
          & \multicolumn{5}{c|}{\textbf{ETM}}
          & \multicolumn{5}{c|}{\textbf{IM}}
          & \multicolumn{5}{c}{\textbf{SM}} \\\cline{2-16}

          \textbf{Measure}
          & \textbf{Fit.}
          & \textbf{Prec.}
          & \textbf{F-score}
          & \textbf{Size}
          & \textbf{CFC}
          & \textbf{Fit.}
          & \textbf{Prec.}
          & \textbf{F-score}
          & \textbf{Size}
          & \textbf{CFC}
          & \textbf{Fit.}
          & \textbf{Prec.}
          & \textbf{F-score}
          & \textbf{Size}
          & \textbf{CFC}  \\\hline

    \textbf{var-e} &  .01(-) &   &   &   &   &   &   &   &   &   &   &   &  .01(-) &   & \\
    \textbf{seq-e} &   &   &   &   &   &   &   &   &   &   &   &   &   &   & \\
    \textbf{nvar-e} &   &   &   &   &   &   &   &   &   &   &   &   &   &   & \\
    \textbf{nseq-e} &  .01(-) &   &   &  .01  &   &   &  .01(-) &  .01(-) &  .05  &  .05  &  .01(-) &  .05(-) &  .01(-) &  .05  & \\
    \textbf{LOD} &   &   &   &  .01  &   &   &  .05(-) &  .05(-) &  .01  &  .01  &  .01(-) &   &  .05(-) &   & \\
    \textbf{granularity} &   &   &   &   &   &   &   &   &   &   &   &   &   &   & \\
    \textbf{structure} &   &   &   &  .01  &   &   &   &   &  .01  &  .01  &   &   &   &  .05  & \\
    \textbf{affinity} &  .05  &   &   &  .01(-) &  .05(-) &   &  .05  &   &  .01(-) &  .01(-) &  .01  &   &  .05  &  .01(-) &  .05(-) \\
    \textbf{dev-random} &   &   &   &  .01  &   &   &  .05(-) &  .05(-) &  .01  &  .01  &  .01(-) &   &  .05(-) &  .01  & \\
    \textbf{avg-dist.} &  .01(-) &   &  .01(-) &  .01  &   &   &  .01(-) &  .01(-) &  .01  &  .01  &  .05(-) &  .01(-) &  .01(-) &  .05  &  .05 \\
    \textbf{LZ} &   &   &   &   &   &   &   &   &   &   &   &   &   &   & \\
    \textbf{t-comp} &  .05(-) &  .05  &   &   &   &   &   &   &   &   &   &   &  .05(-) &   & \\
    \textbf{magnitude} &   &   &   &   &   &   &   &   &   &   &   &   &   &   & \\
    \textbf{support} &   &   &   &   &   &   &   &   &   &   &   &   &   &   & \\
    \textbf{variety} &  .01(-) &   &   &  .01  &   &   &  .05(-) &  .05(-) &  .01  &  .01  &   &   &  .05(-) &  .01  & \\
    \textbf{DT (\%)} &  .01(-) &   &   &  .01  &   &   &  .05(-) &  .05(-) &  .05  &  .05  &  .01(-) &   &  .01(-) &   & \\
    \textbf{DT (\#)} &   &   &   &   &   &   &   &   &   &   &   &   &   &   & \\
    \textbf{TL-min} &   &   &   &   &   &   &   &   &   &   &   &   &   &   & \\
    \textbf{TL-avg} &  .01(-) &   &  .05(-) &  .01  &   &   &  .01(-) &  .01(-) &  .01  &  .01  &  .01(-) &  .05(-) &  .01(-) &  .01  &  .05 \\
    \textbf{TL-max} &  .01(-) &   &  .05(-) &   &   &   &  .05(-) &  .05(-) &   &   &   &  .05(-) &  .01(-) &   & \\
    \hline

        \end{tabular}%
    }}}
  \caption{Overlapping of Pearson and Kendall statically significant correlated measures (at  .01 or  .05), negative correlation is highlighted with the symbol (-)}
  \label{tab:pearken-corr}%
\end{table}%
\endgroup
Focusing on the quality measures of the  automatically discovered process models by ETM (columns 2 to 6, Table~\ref{tab:pearken-corr}), we notice that only one of the log complexity measures correlates with the precision and the CFC of ETM's process models (with a p-value at $0.05$). This highlights that both the precision and the CFC of the process models discovered by ETM are not affected by the complexity of the input log.
A similar observation can be made for the fitness of the process models discovered by IM. By design, IM always strives to discover a highly fitting process model~\cite{leemans2014infrequent}. Consequently, the overall complexity of the input log does not influence the fitness of IM: no correlation exists between the log complexity measures and the fitness of IM's process models.
Finally, also the CFC of the models discovered by SM appears to be resistant to the log complexity, correlating with only three complexity measures: \emph{affinity}, \emph{avg-dist}, and \emph{TL-avg}.

Next, we look at the process model quality measures that correlate the most with the log complexity measures. We find that
the fitness and size of the models discovered by ETM correlate both with 9 log complexity measures;
the precision, F-score, size, and CFC of the models discovered by IM correlate with 8 or 9 log complexity measures;
and the fitness, F-score, and size of the models discovered by SM correlate with 7, 11, and 7 log complexity measures, respectively. We note that when the log complexity measures correlate with the process model accuracy measures (fitness, precision, and their F-score), they almost always (48 out of 52 times) show a negative correlation. There is only one positive correlation recorded for \emph{affinity}, which measures the overlapping directly-follows relations observed in two event sequences. This shows that some features of the process behavior -- captured via the log complexity measures -- may have a negative impact on the accuracy of the automatically discovered process models.

Turning our attention to the log complexity measures, we note that $40\%$ of them (8 out of 20) do not correlate with any process model quality measure. 
These log complexity measures are \emph{seq-e}, \emph{nvar-e}, \emph{granularity}, \emph{lempel-ziv}, \emph{magnitude}, \emph{support}, \emph{DT (\#)}, and \emph{TL-min}.
On the other hand, although, there is no log complexity measure that consistently affects all the quality measures of all the automatically discovered process models, we can identify a subset that correlates with the majority. This subset includes \emph{nseq-e}, \emph{affinity}, \emph{dev-random}, \emph{variety}, \emph{DT (\%)}, \emph{avg-dist}, and \emph{TL-avg}. The latter two correlate with 12 out of 15 quality measures, which makes them the log complexity measures affecting the most the quality of the discovered process models. It is worth noting that while the total number of distinct traces (DT(\#)) does not correlate with any process model quality measure, the percentage of distinct traces (DT(\%)) correlates with more than $50\%$ of them (8 out of 15).

Considering only the measures that correlate the most and selecting only the correlations with a p-value of $0.01$, we are left with the results shown in Table~\ref{tab:corr-subset}.
   \begingroup
\begin{table}[tbp]
\makebox[\textwidth][c]{
\centering
{\scriptsize{
    \begin{tabular}{l|l|l|l|l|l|l|l|l|l}

          \textbf{Log Comp.}
          & \multicolumn{2}{c|}{\textbf{ETM}}
          & \multicolumn{4}{c|}{\textbf{IM}}
          & \multicolumn{3}{c}{\textbf{SM}} \\\cline{2-10}

          \textbf{Measure}
          & \textbf{Fit.}
          & \textbf{Size}
          & \textbf{Prec.}
          & \textbf{F-score}
          & \textbf{Size}
          & \textbf{CFC}
          & \textbf{Fit.}
          & \textbf{F-score}
          & \textbf{Size} \\\hline

    \textbf{nseq-e} &  .01(-) &  .01  &  .01(-) &  .01(-) &       &       &  .01(-) &  .01(-) &  \\
    \textbf{affinity} &       &  .01(-) &       &       &  .01(-) &  .01(-) &  .01  &       &  .01(-) \\
    \textbf{dev-random} &       &  .01  &       &       &  .01  &  .01  &  .01(-) &       &  .01 \\
    \textbf{avg-dist.} &  .01(-) &  .01  &  .01(-) &  .01(-) &  .01  &  .01  &       &  .01(-) &  \\
    \textbf{variety} &  .01(-) &  .01  &       &       &  .01  &  .01  &       &       &  .01 \\
    \textbf{DT (\%)} &  .01(-) &  .01  &       &       &       &       &  .01(-) &  .01(-) &  \\
    \textbf{TL-avg} &  .01(-) &  .01  &  .01(-) &  .01(-) &  .01  &  .01  &  .01(-) &  .01(-) &  .01 \\\hline

        \end{tabular}%
    }}}
  \caption{Log complexity measures and process model quality measures that correlate the most, with emphasis on correlations with $p < 0.01$}
  \label{tab:corr-subset}%
\end{table}%
\endgroup

\subsection{Results of Regression Analysis}
We now want to investigate to what extent log complexity measures could be used to predict a process model quality measure. 
To this end, for each log complexity measure that correlates with a process model quality measure in Table~\ref{tab:corr-subset}, we assessed the \emph{residual errors} (minimum, median, and maximum) and the coefficient of determination (R\textsuperscript{2}) of the corresponding linear regression model, where the log complexity measure is the \emph{predictor variable} and the process model quality measure is the \emph{outcome variable}.
The linear regression models that exhibit the highest R\textsuperscript{2} should be preferred, since R\textsuperscript{2} is defined as $1-\frac{SS_r}{SS_t}$, where $SS_r$ is \emph{the residual sum of squares} and $SS_t$ is \emph{the total sum of squares}. In general, the higher R\textsuperscript{2} the more accurate is the linear regression model.
The results of this analysis are reported in Table~\ref{tab:reg-models}.
Looking at the process model accuracy measures (fitness\textsubscript{ETM}, precision\textsubscript{IM}, F-score\textsubscript{IM}, fitness\textsubscript{SM}, and F-score\textsubscript{SM}), we note that the best complexity measures to predict them are \emph{avg-dist} (for ETM and IM) and \emph{nseq-e} (for SM). This shows that the accuracy of the process models discovered by ETM, IM, and SM is closely affected by the variation of process behavior recorded in the event logs, rather than the absolute amount of process behavior. Therefore, automatically discovering a process model is likely to be more challenging when the event log records a small amount of process behavior that varies greatly than when the event log records a huge amount of process behavior that varies little.
Looking at the model complexity measures Size\textsubscript{ETM}, CFC\textsubscript{ETM}, CFC\textsubscript{IM}, and Size\textsubscript{SM}, one log complexity measure seems to highly affect them all, which is \emph{variety}. 

It is possible to generate more accurate regression models by using \emph{uncorrelated} multi-predictors or non-linear regression models (e.g., generalized additive models).
Ideally, having an accurate regression model to estimate the quality of the automatically discovered process models could save considerable time for process analysts, especially if an optimization is required during process discovery~\cite{augusto2018automated,augusto2019metaheuristic}. In fact, one could select the automated process discovery algorithm to be used based on the results of the regression model predictions. However, the design of such an accurate regression model and its systematic evaluation would deserve a separate study, a very different dataset (e.g., a large collection of event logs, even artificial), and a different type of evaluation, as seen in previous studies~\cite{mendling2012thresholds}. We also note that seminal studies in that area have been conducted by Riberio et al.~\cite{ribeiro2014recommender,ribeiro2015method}, however, 
neither taking into account the log complexity measures nor the novel automated process discovery algorithms designed in the past five years (including, Fodina~\cite{vanden2017fodina} and Split Miner~\cite{augusto2018split}). Furthermore, Riberio et al. propose a black-box prediction system, which does not focus on the connection between log complexity measures and the quality of the discovered process model.

\begingroup
\begin{table}[tbp]
\makebox[\textwidth][c]{
\centering
{\scriptsize{
    \begin{tabular}{l|l|rrr|rr|r}

    \textbf{Outcome }
    & \textbf{Predictor}
    & \multicolumn{3}{c|}{\textbf{Residual Errors}}
    & \multicolumn{2}{c|}{\textbf{Predictor}}
    &  \\\cline{3-7}

    \textbf{Variable}
    & \textbf{Variable}
    & \multicolumn{1}{c}{\textbf{min}}
    & \multicolumn{1}{c}{\textbf{median}}
    & \multicolumn{1}{c|}{\textbf{max}}
    & \multicolumn{1}{c}{\textbf{coeff.}}
    & \multicolumn{1}{c|}{\textbf{p-value}}
    & \multicolumn{1}{c}{\textbf{R\textsuperscript{2}}}\\\hline

              & \textbf{nseq-e} & -0.368 & \textbf{0.005} & 0.333 & -0.747 & 0.003 & 0.275 \\
          & \textbf{avg-dist.} & \textbf{-0.284} & 0.014 & \textbf{0.135} & -0.013 & \textbf{0.000} & \textbf{0.632} \\
    \textbf{ETM Fitness} & \textbf{variety} & -0.533 & 0.040 & 0.223 & -0.004 & 0.008 & 0.223 \\
          & \textbf{DT (\%)} & -0.530 & 0.022 & 0.229 & -0.003 & 0.007 & 0.234 \\
          & \textbf{TL-avg} & -0.398 & 0.024 & 0.180 & -0.011 & \textbf{0.000} & 0.435 \\\hline

          & \textbf{nseq-e} & -38.617 & \textbf{1.017} & 29.276 & 87.706 & 0.003 & 0.282 \\
          & \textbf{affinity} & -44.328 & 1.087 & \textbf{23.366} & -93.835 & \textbf{0.000} & 0.463 \\
          & \textbf{dev-random} & -38.982 & 2.470 & 27.335 & 126.467 & \textbf{0.000} & 0.507 \\
    \textbf{ETM Size} & \textbf{avg-dist.} & -28.108 & 2.151 & 26.790 & 1.212 & \textbf{0.000} & 0.410 \\
          & \textbf{variety} & \textbf{-20.198} & -2.628 & 29.969 & 0.767 & \textbf{0.000} & \textbf{0.658} \\
          & \textbf{DT (\%)} & -39.073 & -1.568 & 31.732 & 0.408 & 0.006 & 0.239 \\
          & \textbf{TL-avg} & -38.768 & 1.606 & 31.796 & 1.268 & \textbf{0.000} & 0.432 \\\hline

          & \textbf{nseq-e} & -0.383 & 0.030 & 0.331 & -0.841 & 0.001 & 0.351 \\
    \textbf{IM Precision} & \textbf{avg-dist.} & -0.344 & 0.041 & \textbf{0.280} & -0.011 & \textbf{0.000} & \textbf{0.423} \\
          & \textbf{TL-avg} & \textbf{-0.330} & \textbf{0.012} & 0.292 & -0.010 & 0.001 & 0.331 \\\hline

          & \textbf{nseq-e} & -0.387 & 0.042 & 0.195 & -0.636 & 0.001 & 0.316 \\
    \textbf{IM F-score} & \textbf{avg-dist.} & -0.318 & 0.047 & \textbf{0.177} & -0.008 & \textbf{0.000} & \textbf{0.401} \\
          & \textbf{TL-avg} & \textbf{-0.316} & \textbf{0.028} & 0.219 & -0.007 & 0.002 & 0.300 \\\hline

          & \textbf{affinity} & -67.971 & -7.348 & 105.713 & -177.023 & 0.002 & 0.295 \\
          & \textbf{dev-random} & -67.632 & 2.120 & 75.020 & 317.187 & \textbf{0.000} & 0.570 \\
    \textbf{IM Size} & \textbf{avg-dist.} & -80.604 & -8.423 & 74.846 & 3.090 & \textbf{0.000} & 0.477 \\
          & \textbf{variety} & \textbf{-31.157} & \textbf{0.109} & \textbf{20.143} & 2.192 & \textbf{0.000} & \textbf{0.963} \\
          & \textbf{TL-avg} & -93.232 & -4.092 & 67.909 & 3.571 & \textbf{0.000} & 0.614 \\\hline

          & \textbf{affinity} & -48.798 & -4.414 & 64.821 & -130.587 & 0.001 & 0.327 \\
          & \textbf{dev-random} & -46.816 & \textbf{0.205} & 44.763 & 220.020 & \textbf{0.000} & 0.558 \\
    \textbf{IM CFC} & \textbf{avg-dist.} & -57.225 & -6.559 & 47.295 & 2.353 & \textbf{0.000} & 0.562 \\
          & \textbf{variety} & \textbf{-16.869} & -1.123 & \textbf{14.600} & 1.539 & \textbf{0.000} & \textbf{0.965} \\
          & \textbf{TL-avg} & -67.689 & -3.600 & 43.697 & 2.598 & \textbf{0.000} & 0.661 \\\hline

          & \textbf{nseq-e} & -0.147 & 0.010 & 0.132 & -0.387 & \textbf{0.000} & \textbf{0.405} \\
          & \textbf{affinity} & -0.144 & \textbf{0.008} & 0.125 & 0.284 & 0.001 & 0.313 \\
    \textbf{SM Fitness} & \textbf{dev-random} & -0.141 & 0.023 & \textbf{0.102} & -0.361 & 0.002 & 0.305 \\
          & \textbf{DT (\%)} & \textbf{-0.127} & 0.013 & 0.194 & -0.002 & 0.001 & 0.308 \\
          & \textbf{TL-avg} & -0.155 & 0.018 & 0.151 & -0.003 & 0.007 & 0.231 \\\hline

          & \textbf{nseq-e} & \textbf{-0.137} & -0.002 & 0.089 & -0.357 & \textbf{0.000} & \textbf{0.511} \\
    \textbf{SM F-score} & \textbf{avg-dist.} & \textbf{-0.137} & 0.014 & 0.071 & -0.004 & \textbf{0.000} & 0.497 \\
          & \textbf{DT (\%)} & -0.221 & \textbf{0.001} & \textbf{0.070} & -0.001 & 0.001 & 0.315 \\
          & \textbf{TL-avg} & -0.170 & 0.015 & 0.074 & -0.004 & \textbf{0.000} & 0.438 \\\hline

          & \textbf{affinity} & -93.412 & \textbf{-0.441} & 295.530 & -226.293 & 0.006 & 0.243 \\
    \textbf{SM Size} & \textbf{dev-random} & -77.411 & -11.069 & 311.668 & 279.926 & 0.008 & 0.224 \\
          & \textbf{variety} & \textbf{-54.125} & -9.066 & 284.199 & 2.017 & \textbf{0.000} & \textbf{0.411} \\
          & \textbf{TL-avg} & -108.224 & -9.111 & \textbf{239.024} & 3.965 & \textbf{0.000} & 0.381 \\\hline

    \end{tabular}%
    }}}
  \caption{Results of the regression analysis}\label{tab:reg-models}%
\end{table}%
\endgroup

\subsection{Results of Computational Performance}
Finally, we compared the computational performance of all the log complexity measures, assessing their average, maximum, minimum, and median execution times over the 32 event logs. The results are shown in Table~\ref{tab:performance}. We did not report the execution times of derived measures (e.g., the percentage of distinct traces, which is derived from the total number of distinct traces and support), or of ratios (e.g., our normalized graph entropy measures).
We note that all the complexity measures have a median execution time below one second, with the exception of \emph{affinity} with a median execution time of $9.6$ seconds. Indeed, \emph{affinity} was the slowest measure to be calculated, followed by our graph entropy-based complexity measures and \emph{avg-dist}. However, while \emph{affinity} has an average execution time well above a minute, our measures and \emph{avg-dist} have an average execution time in the order of seconds, which is reasonable given that these measures are not designed to work in a real-time context.

\begingroup
\begin{table}[tbp]
\makebox[\textwidth][c]{
\centering
{\scriptsize{
    \begin{tabular}{l|c|c|c|c|c|c|c}
          & \textbf{var-e}
          & \textbf{seq-e}
          & \textbf{LOD}
          & \textbf{granularity}
          & \textbf{structure}
          & \textbf{affinity}
          & \textbf{dev-random} \\\hline
    \textbf{AVG} & 22.1  & 21.7  & 0.2   & 0.2   & 0.6   & 148.4 & 0.8 \\
    \textbf{MAX} & 397.5 & 384.2 & 2.2   & 1.8   & 5.8   & 2020.1 & 10.7 \\
    \textbf{MIN} &  $<0.1$   &  $<0.1$   &  $<0.1$   &  $<0.1$   &  $<0.1$   &  $<0.1$   &  $<0.1$ \\
    \textbf{MEDIAN} & 0.1   & 0.1   &  $<0.1$   &  $<0.1$   & 0.2   & 9.6   & 0.1 \\\hline
    \end{tabular}%
}}}

\medskip

\makebox[\textwidth][c]{
\centering
{\scriptsize{
    \begin{tabular}{l|c|c|c|c|c|c|c|c}
    \hline
          & \textbf{avg-dist.}
          & \textbf{LZ}
          & \textbf{t-comp}
          & \textbf{magnitude}
          & \textbf{support}
          & \textbf{variety}
          & \textbf{DT (\%)}
          & \textbf{TL-min} \\\hline
    \textbf{AVG} & 18.1  & 0.6   &  $<0.1$   &  $<0.1$   &  $<0.1$   & 0.2   & 0.2   & 0.1 \\
    \textbf{MAX} & 176.0 & 8.4   &  $<0.1$   &  $<0.1$   &  $<0.1$   & 2.0   & 1.8   & 0.9 \\
    \textbf{MIN} &  $<0.1$   &  $<0.1$   &  $<0.1$   &  $<0.1$   &  $<0.1$   &  $<0.1$   &  $<0.1$   &  $<0.1$ \\
    \textbf{MEDIAN} & 0.6   &  $<0.1$   &  $<0.1$   &  $<0.1$   &  $<0.1$   &  $<0.1$   &  $<0.1$   &  $<0.1$ \\\hline
    \end{tabular}%
    }}}
  \caption{Execution times (in seconds) of the complexity measures}
  \label{tab:performance}%
\end{table}%
\endgroup
\subsection{Threats to Validity}

The findings reported in this study should be interpreted taking into account the size and variety of the 32 event logs of our dataset. Although the event logs are a good approximation of event logs that one may find in real-world scenarios, both in terms of complexity and variety of domains, 
they cannot possibly synthesize all possible business processes observable in an industrial setting. It is interesting to note that the original submission of this study's manuscript included only 24 event logs~\cite{augusto2019dataset} and that the additional 8 event logs were added following the reviewers' suggestions. Nonetheless, the changes in the statistical analysis were minor. For instance, the total number of pairs of measures that correlated according to both Pearson and Kendall correlations increased from 84 to 90 (see Table~\ref{tab:pearken-corr}). Even more importantly, the seven log complexity measures that correlated the most did not change (Table~\ref{tab:corr-subset}), nor did the most robust predictors (Table~\ref{tab:reg-models}). 

Another threat to the validity of our evaluation is the selection of the quality measures of automatically discovered process models, which in the past few years have been under scrutiny~\cite{Tax20181,syring11790evaluating} in the process mining research community.
However, these research studies showed that no existing precision measure is ideal, while newly designed ones are either approximate~\cite{augusto2020accuracy} or computationally inefficient~\cite{8786053,polyvyanyy2020monotone}. Although one may argue that one precision measure is better than another, we note that the choice of the precision measure does not affect the results much. The findings are reliable and accurate in light of the quality measures we used, which remain the most popular as of today.


\section{Conclusions}~\label{sec:conclude}
With this paper, we provide two major contributions to measuring process complexity and to assessing process mining algorithms.
First, we analyzed existing measures for process complexity that are based on event logs. Each of these measures emphasizes different complexity criteria including size, variation, and distance. We defined new measures of process complexity based on graph entropy, which capture all three concerns of process complexity by adhering to monotonicity.
Second, we evaluated the identified set of process complexity measures, including our novel measures, using a benchmark collection of event logs and their corresponding automatically discovered process models. The goal of our evaluation was to investigate which empirical connections hold between the process complexity measures and the quality of discovered models. Our results show that many process complexity measures (including our novel measure) correlate with the quality of the discovered process models and that it is possible to use process complexity measures as predictors for the quality of process models discovered with state-of-the-art process discovery algorithms.

The findings we reported in this paper are important for process mining research, as they highlight that not only algorithms but also empirical connections between input data complexity and output quality should be investigated. Our results demonstrate the potential to examine the concept of process complexity and its corresponding measures in connection with automated process discovery and there are various opportunities to extend this approach to related research problems. Additional aspects of event log data, such as data complexity, could be used to study connections with further output parameters, such as the process model's usefulness perceived by analysts. 

\section*{Acknowledgement}
\noindent Research by Jan Mendling is funded by the Einstein Foundation Berlin under grant number EPP-2019-524.

\bibliography{mybibfile}

\end{document}